\pdfoutput=1

\documentclass[reqno,a4paper]{article}
\usepackage{amssymb,amsmath,epsfig,float,psfrag,fancyhdr,color,subfig}
\definecolor{lightblue}{rgb}{0.22,0.45,0.70}
\usepackage[square,sort,comma,numbers]{natbib}

\numberwithin{equation}{section}
\numberwithin{table}{section}
\numberwithin{figure}{section}

\usepackage[colorlinks=true,breaklinks=true,linkcolor=lightblue,citecolor=lightblue]{hyperref}

\newcommand{\mbs}[1]{\mathbf{#1}}

\definecolor{lightblue}{rgb}{0.22,0.45,0.70}

    \def\bC{{\mbs{C}}}
    \def\bF{{\mbs{F}}}
  \def\bH{{\mbs{H}}}  \def\bI{{\mbs{I}}}

\def\bS{{\mbs{S}}}    
    
  \def\b0{{\mbs{0}}}

\def\cT{{\mathcal{T}}}

\newcommand\bphi{\boldsymbol{\varphi}}
\newcommand\bpsi{\boldsymbol{\psi}}

\def\det{{\rm det}\,}
\def\tr{{\rm tr}\,}

\def\Isa	{{{\rm I}_{\rm sac} {(\lambda,u)}}}
\def\usa	{{u_{\rm sac} }}

\allowdisplaybreaks

\begin{document}
\title{Competing mechanisms of stress-assisted diffusivity and stretch-activated currents in cardiac electromechanics}
\author{{\sc Alessandro Loppini}\thanks{Nonlinear Physics and Mathematical Modeling, Department of Engineering, University Campus Bio-Medico, Rome, Italy. E-mail: {\tt \{a.loppini,a.gizzi,c.cherubini,s.filippi\}@unicampus.it}.}\ , \quad 
{\sc Alessio Gizzi$^{*}$}, \quad 
{\sc Ricardo Ruiz Baier}\thanks{Mathematical Institute,
University of Oxford, A. Wiles Building, 
Woodstock Road, Oxford OX2 6GG, UK. E-mail: {\tt ruizbaier@maths.ox.ac.uk}.}\ ,\\
{\sc Christian Cherubini$^{*,}$}\thanks{International Center for Relativistic Astrophysics (ICRA), University Campus Bio-Medico, Rome, Italy; and ICRANet, Piazza delle Repubblica 10, I-65122 Pescara, Italy.}\ , \quad
{\sc Flavio H. Fenton}\thanks{Georgia Institute of Technology, School of Physics, 837 State Street, Atlanta, Georgia, USA.}\ , \quad     
{\sc Simonetta Filippi$^{*,\ddag}$}}

\date{December 18, 2017}
\maketitle


\begin{abstract}
\noindent We numerically investigate the role of mechanical stress in modifying the conductivity properties of the cardiac tissue and its impact in computational models for cardiac electromechanics. We follow a theoretical framework recently proposed in \cite{cherubini:2017}, in the context of general reaction-diffusion-mechanics systems using multiphysics continuum mechanics and finite elasticity. In the present study, the adapted models are compared against preliminary experimental data of pig right ventricle fluorescence optical mapping. These data contribute to the characterization of the observed inhomogeneity and anisotropy properties that result from mechanical deformation. Our novel approach simultaneously incorporates two mechanisms for mechano-electric feedback (MEF): stretch-activated currents (SAC) and stress-assisted diffusion (SAD); and we also identify their influence into the nonlinear spatiotemporal dynamics. It is found that i) only specific combinations of the two MEF effects allow proper conduction velocity measurement; ii) expected heterogeneities and anisotropies are obtained via the novel stress-assisted diffusion mechanisms; iii) spiral wave meandering and drifting is highly mediated by the applied mechanical loading. We provide an analysis of the intrinsic structure of the nonlinear coupling using computational tests, conducted using a finite element method. In particular, we compare static and dynamic deformation regimes in the onset of cardiac arrhythmias and address other potential biomedical applications.
\end{abstract}


\section{Introduction}

Cardiac tissue is a complex multiscale medium constituted by
highly interconnected units, cardiomyocytes, that conform a so-called
\emph{syncitium} with unique structural and functional
properties~\citep{pullan:2005}. Cardiomyocytes are excitable and
deformable muscular cells that present an additional multiscale
architecture in which plasma membrane proteins and intracellular
organelles all depend on the current mechanical state of the
tissue~\citep{salameh:2013,schonleitner:2017}. Dedicated proteic
structures, such as ion channels or gap junctions, rule the passage of
charged particles throughout the cell as well as between different
cells and they are usually described mathematically through multiple
 reaction-diffusion (RD)
systems~\citep{kleber:2014,cabo:2014,dhein:2014}. All these coupled
nonlinear and stochastic dynamics, emerge then to conform the
coordinated contraction and pumping of the
heart~\citep{augustin:2016,land:2015,quarteroni:2017}. During the overall cycle, the
mechanical deformation undoubtedly affects the electrical impulses that
modulate muscle contraction, also modifying the properties of
the substrate where the electrical wave propagates. These multiscale
interactions have commonly been referred in the literature as the
mechano-electric feedback (MEF)~\citep{ravelli:2003}. Experimental,
theoretical and clinical studies have been contributing to the
systematic investigation of MEF effects, already for over a century;
however, several open questions still
remain~\citep{quinn:2014,quinn:2016,land:2017}. 
For example, and focusing on the cellular level, 
it is still now not completely understood what is the effective contribution of 
stretch-activated ion channels and which is the most appropriate way to describe 
them. 
In addition, and focusing on the organ scale, 
the clinical relevance of MEF in patients with heart diseases 
remains an open issue~\citep{taggart:2017}, more specifically, 
how MEF mechanisms translate into ECGs~\citep{MEIJBORG2017356}
and what is the specific role of mechanics during cardiac arrhythmias~
\citep{christoph:2018} is still under investigation.

The theoretical and
computational modeling of cardiac electromechanics has been 
used to investigate some key aspects of general excitation-contraction 
mechanisms. For instance, the
transition from cardiac arrhythmias to chaotic behavior,
including the onset, drift and breakup of spiral/scroll waves
\citep{bini:2010,hunter:1997,trayanova:2006,panfilov:2005,panfilov:2010,trayanova:2010,dierckx:2015,christoph:2018},
pinning and unpinning phenomena due to anatomical
obstacles~\citep{cherubini:2012,horning:2012,chen:2014}, as well as the
multiscale and stochastic dynamics both at subcellular, cellular and
tissue scale~\citep{trayanova:2004,trayanova:2011a,hurtado:2016,land:2017}.
However, the formulation of MEF effects into mathematical
models has been primarily focused on accounting for the additive
superposition of an active and passive stress to stretch-activated
currents~\citep{panfilov:2005}. Recent contributions have advanced an
energy-based framework for the comparison of active stress,
stretch-activated currents and inertia
effects~\citep{ambrosi:2012,hurtado:2017a,cherubini:2008,rossi:2014}. 
These works further highlight the role of mechanics into the resulting heart
function at different temporal and spatial scales.

\begin{figure}[t!]
\begin{center}
	\includegraphics[width=0.75\textwidth]{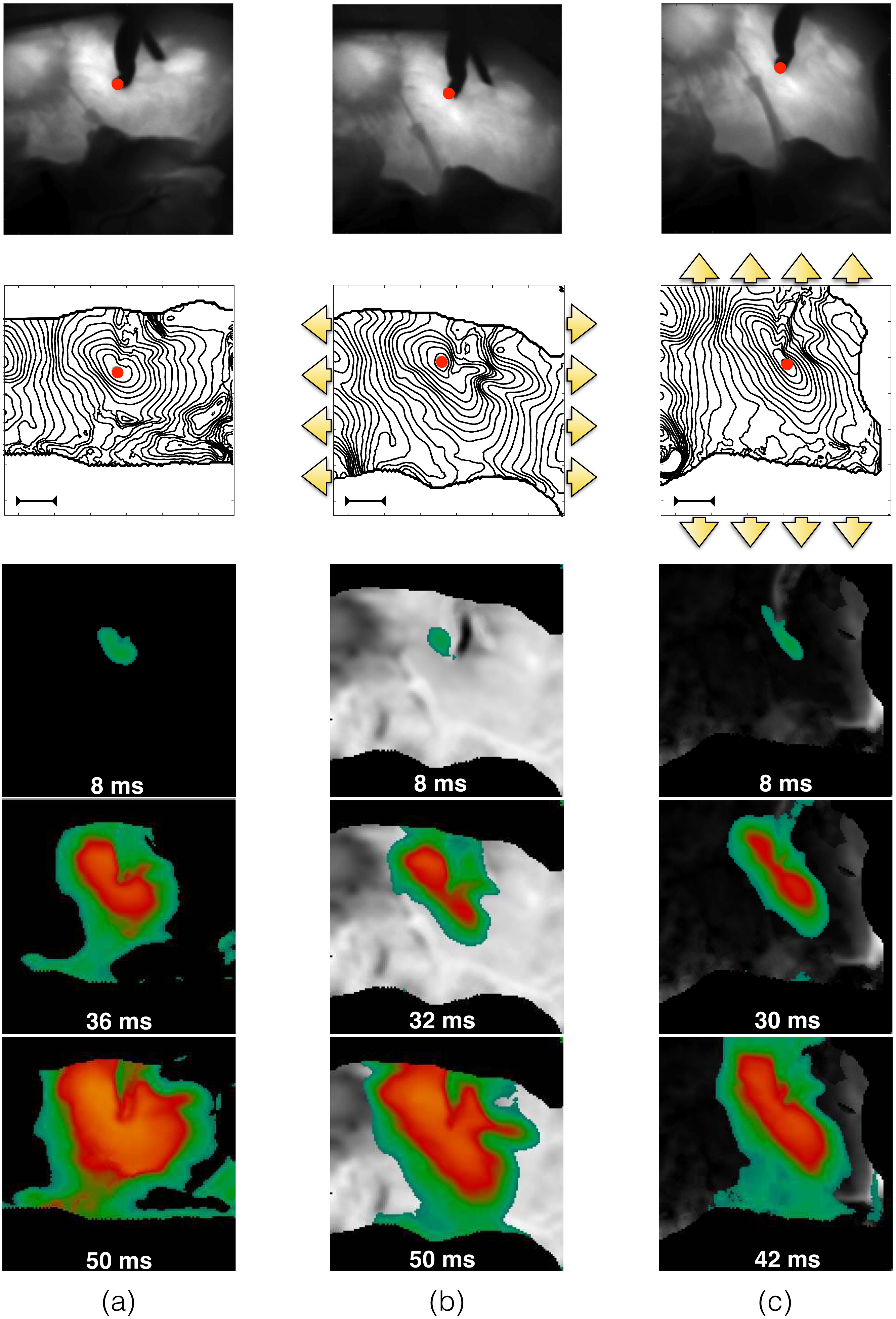}
\end{center}
\caption{MEF observed in pig right ventricle via fluorescence optical mapping.
From top to bottom, we provide:
underlying tissue structure in reference (a) and deformed (b,c) states;
activation isochrones each $4\,{\rm ms}$
originating from the stimulation point (red spot in the field of view--length bar $1\,{\rm cm}$),
and activation sequences.
The three cases refer to no-stretch (a), static horizontally (b), and vertical (c) stretch 
in the directions indicated by the yellow arrows.
The sequence of spatial activation uses
the color code scaled to the AP level (yellow/green -- high/low). 
Selected frames highlight the anisotropy induced by stretch.
The outer black region is the noisy area not useful for the field of view.}
\label{fig:0}
\end{figure}

To further motivate our theoretical considerations we provide
an experimental representative example of the strong MEF coupling in cardiac tissue at the macroscale.
The data shown in Fig.~\ref{fig:0} were obtained via dedicated 
fluorescence optical mapping analyses of a pig right ventricle (the
experimental procedure has been previously described in
\citealp{fenton:2009,gizzi:2013,fenton:2017a}). 
After motion suppression via blebbistatin, 
the perfused tissue was electrically stimulated via an external bipolar stimulator
with strength twice diastolic threshold.
{An excitation pulse with constant pacing cycle length of $1\,{\rm s}$}
was delivered within the field of view (red spot in Fig.~\ref{fig:0}) 
{for several seconds (reaching a steady-state configuration) and}
for three different mechanical loading conditions on the same wedge: 
(a) free edges,
(b) static uniaxial horizontal stretch,
(c) static uniaxial vertical stretch
with respect to a prescribed tissue orientation.
The figure displays
the underlying structure with clear evidence of the deformed tissue architecture,
isochrones of electrical activation {for a representative stimulus},
and a sequence of spatial activation maps, where the
colors indicate the level of activation--Action Potential (AP). 
Since in this proof of concept setup active contraction is inhibited by blebbistatin,
these experiments clearly indicate that an additional degree of heterogeneity and anisotropy
appears in the tissue and affects the AP excitation wave
due to the intensity and direction of the
externally applied deformation. In addition, this behavior does
not correspond to a mere linear mapping from the reference to the
deformed configuration (as a visual scaling of the image would easily
show), but one observes that mechanical deformations induce higher,
nonlinear and non-trivial anisotropies and heterogeneities in the
tissue.

\begin{figure}[htp!]
\begin{center}
	\includegraphics[width=0.325\textwidth]{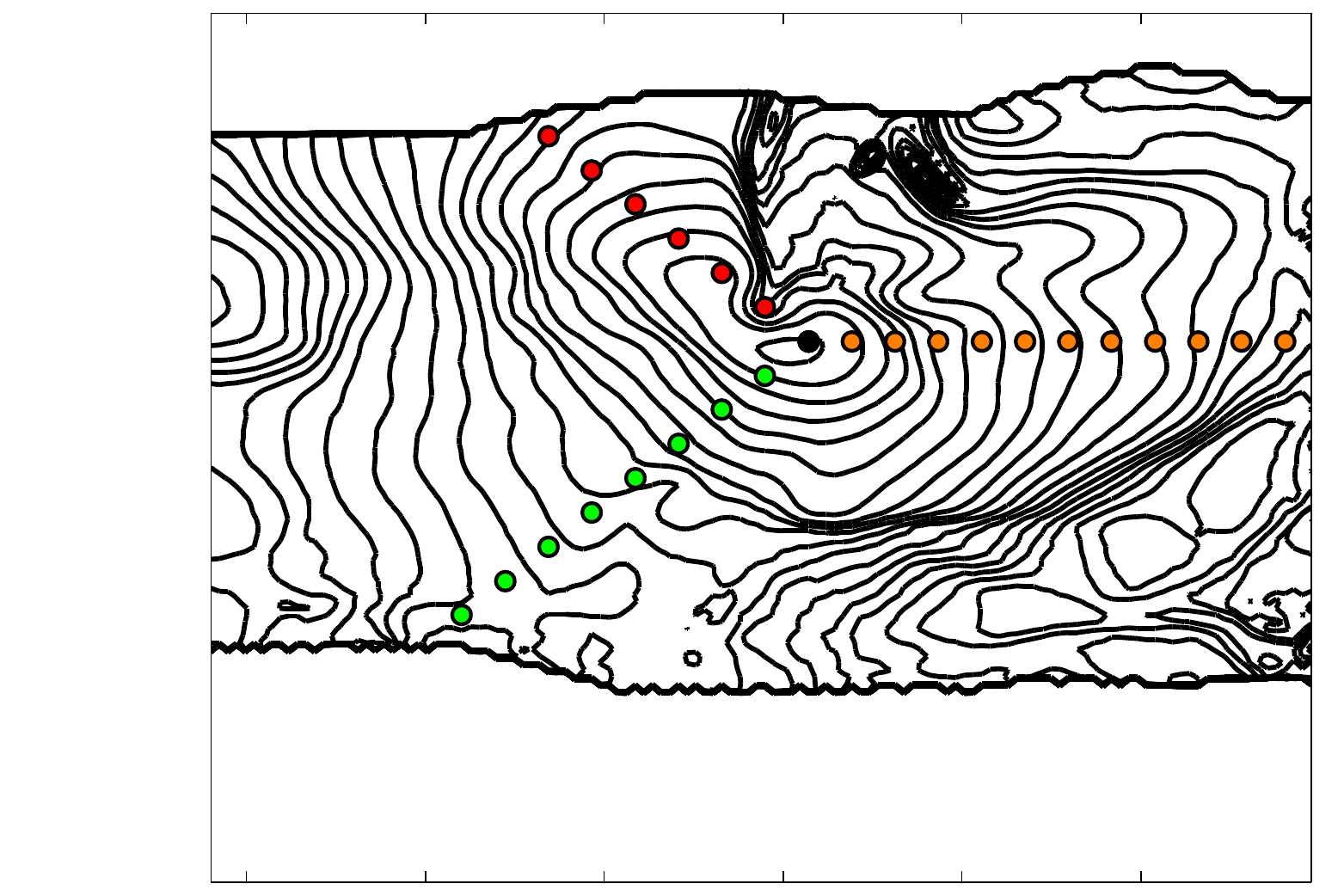}
	\includegraphics[width=0.325\textwidth]{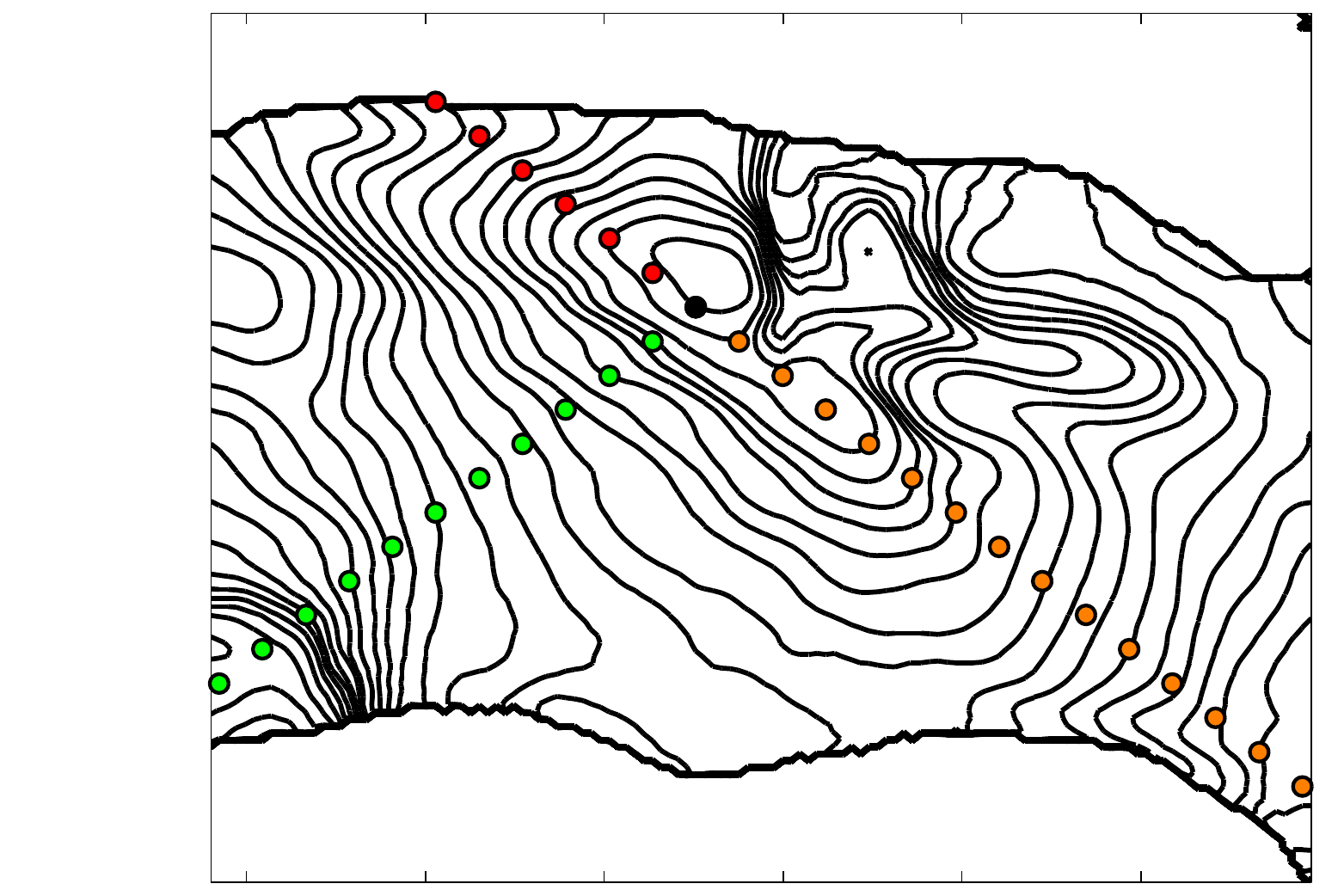}
	\includegraphics[width=0.325\textwidth]{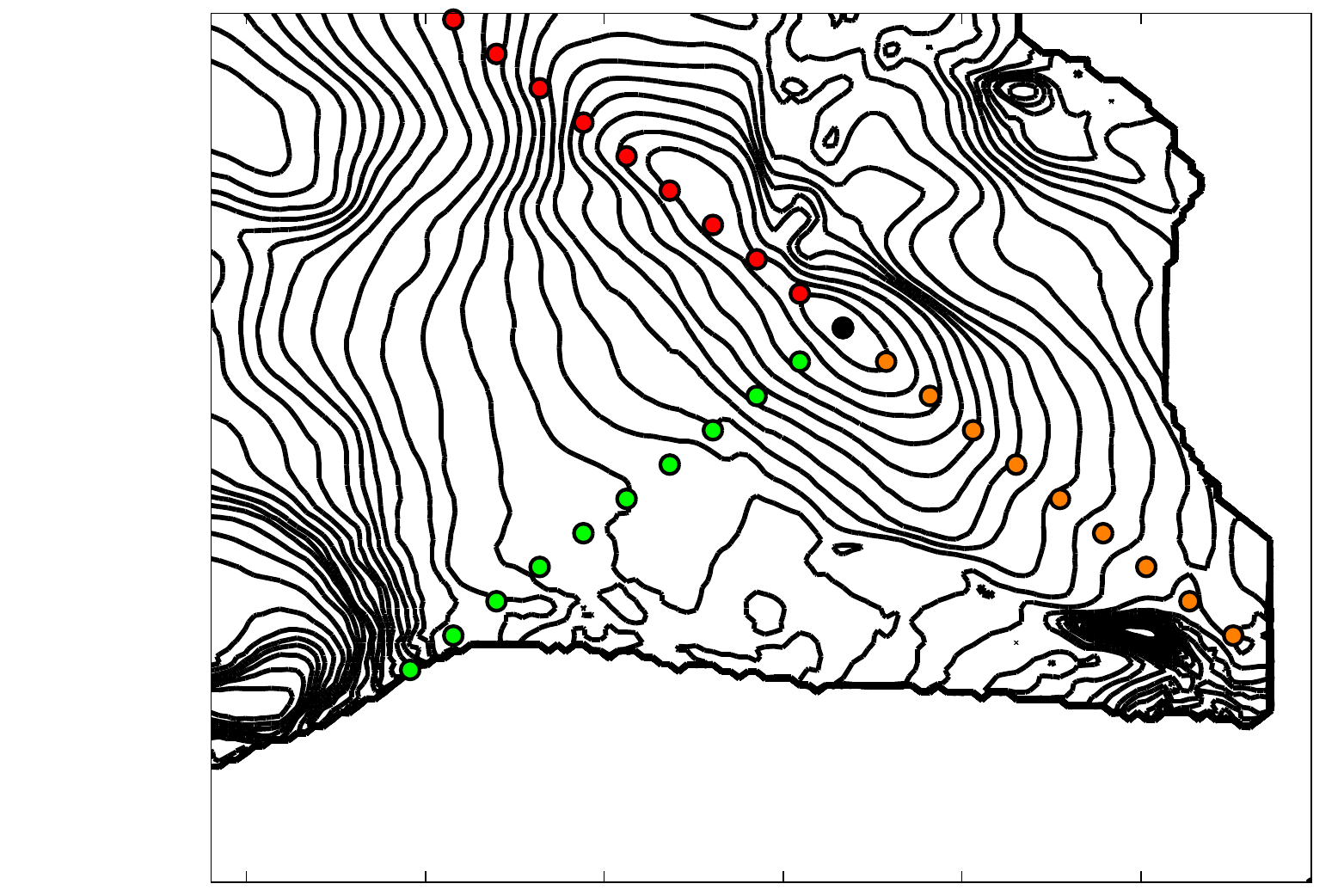}
	\\
	\includegraphics[width=0.325\textwidth]{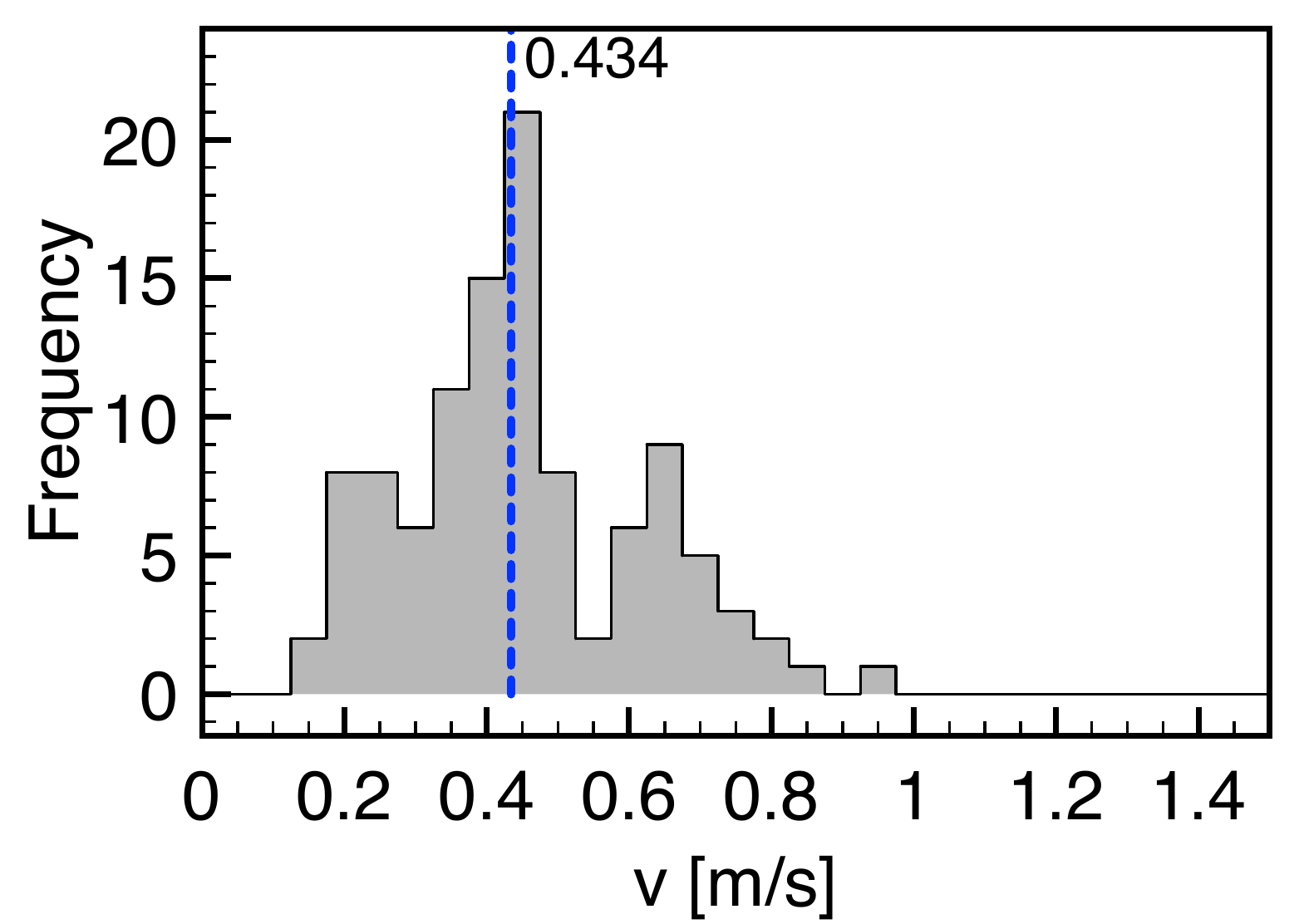}
	\includegraphics[width=0.325\textwidth]{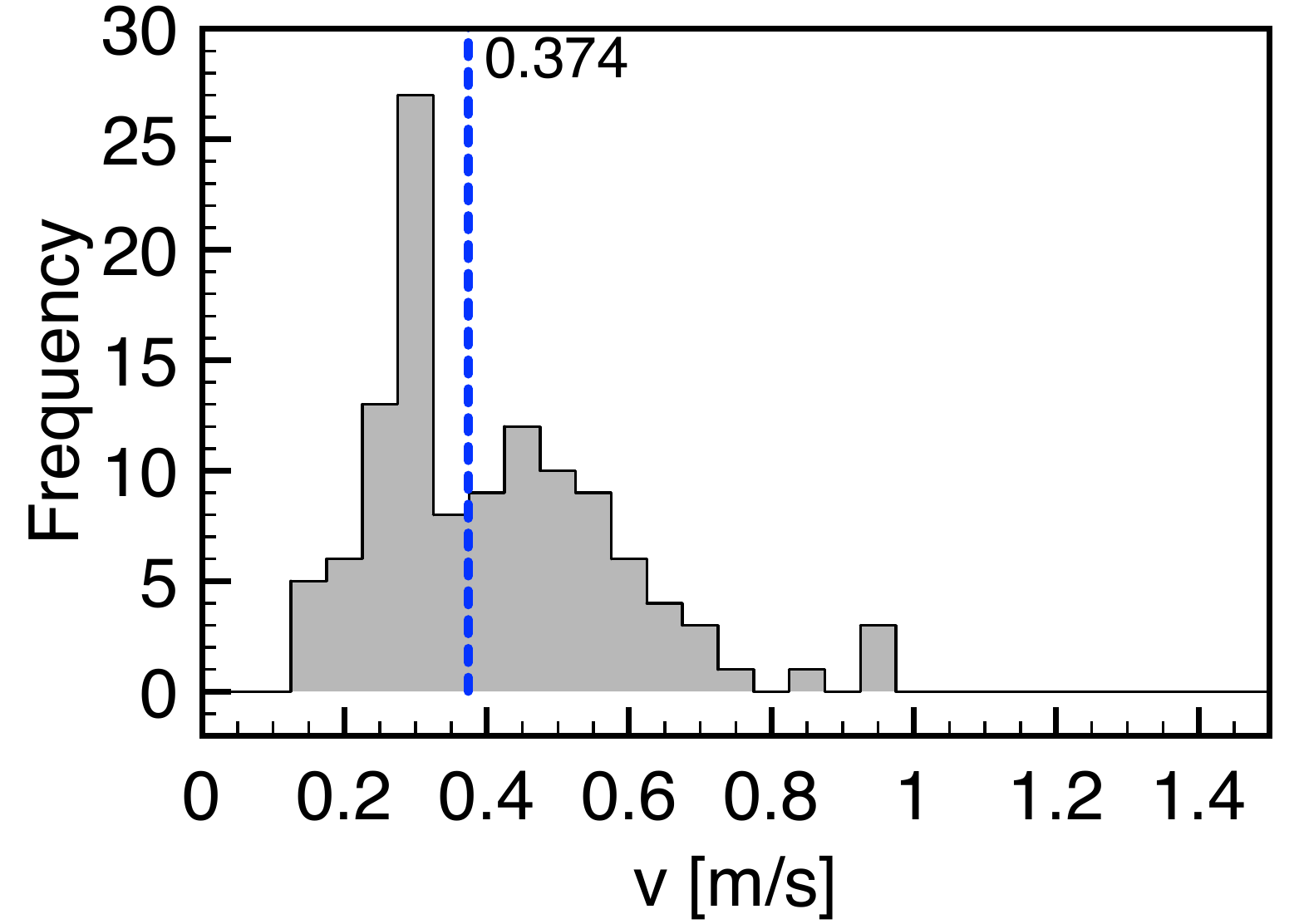}
	\includegraphics[width=0.325\textwidth]{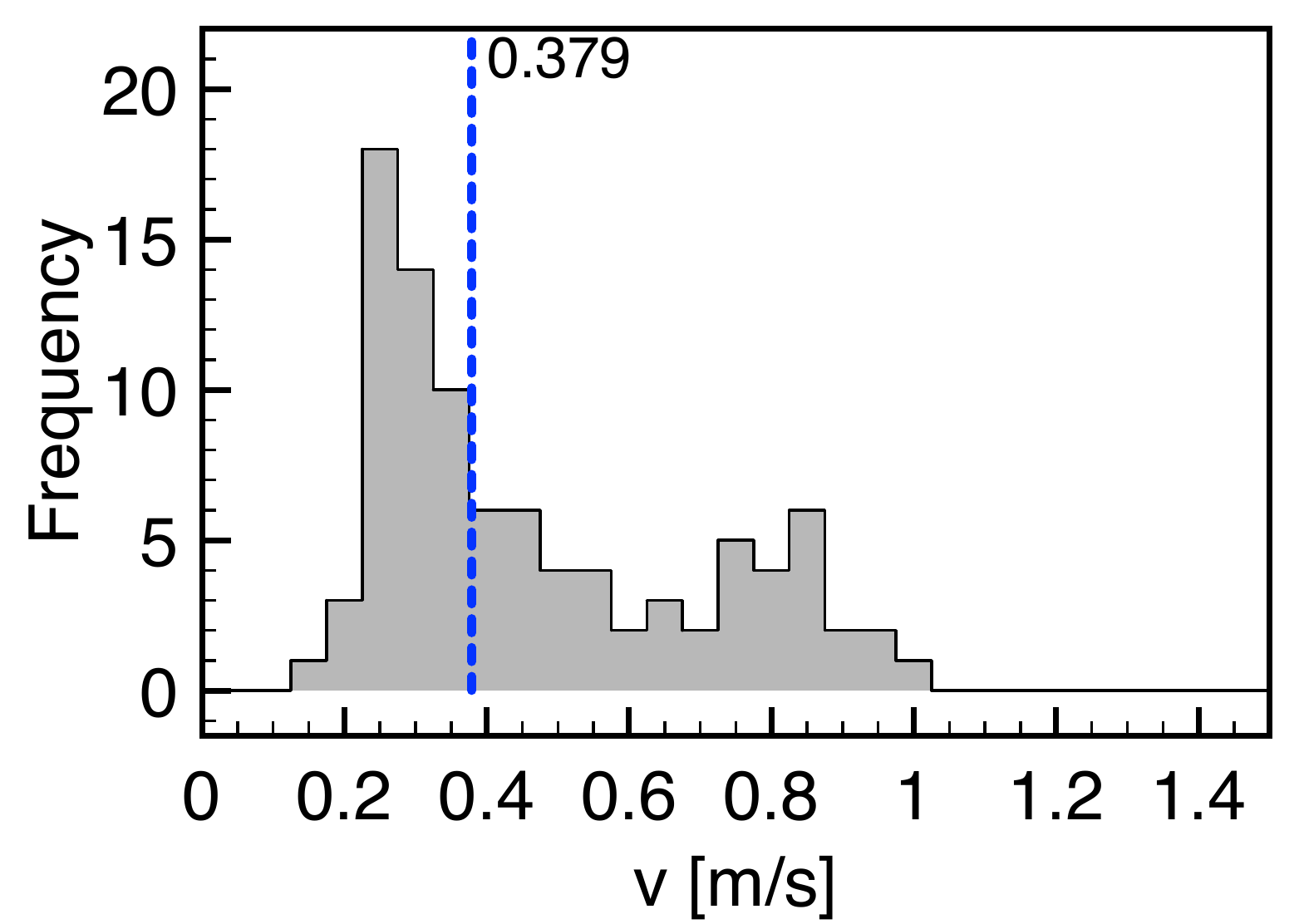}
\end{center}
\caption{CV histograms measured on tissue wedges for three different loading states
{overlapping local measures for five consecutive activations at constant pacing 
cycle length of $1\,{\rm s}$}.
Multiple directions of AP propagation are considered as indicated in the green, 
orange, and red circles seen on the top panels. The median is highlighted 
denoting a reduction of CV under stretch passing from $\sim0.43$ (left) to $\sim0.37\, {\rm m/s}$ in the stretched cases (center, right), respectively (see Fig.~\ref{fig:0}).}
\label{fig:stat}
\end{figure}

{To better characterize such features, in Fig.~\ref{fig:stat} we provide
the histograms of the conduction velocity (CV) measured as follows. 
\begin{itemize}
\item locally on the tissue with a fixed spacing step, 
such to minimize and homogenize tissue heterogeneity,
\item considering multiple directions of propagation (as enhanced on the isochrons panels),
in order to minimize curvature effects of the activation front due to the underlying ventricular structure,
and 
\item overlapping five consecutive activations at constant pacing cycle length of $1\,\rm{s}$,
with the aim to minimize physiological beat-to-beat variabilities.
\end{itemize}
We provide such an extended CV analysis 
for the three loading cases as described in Fig.~\ref{fig:0}.}
According to previous studies~\citep{ravelli:2003}, we proceed to identify
a reduction of the CV median when the tissue undergoes stretching. We will  
regard these velocity values as the reference case, when addressing 
the construction of the proposed model described in what follows.

\begin{figure}[t!]
\begin{center}
	\includegraphics[width=0.86\textwidth]{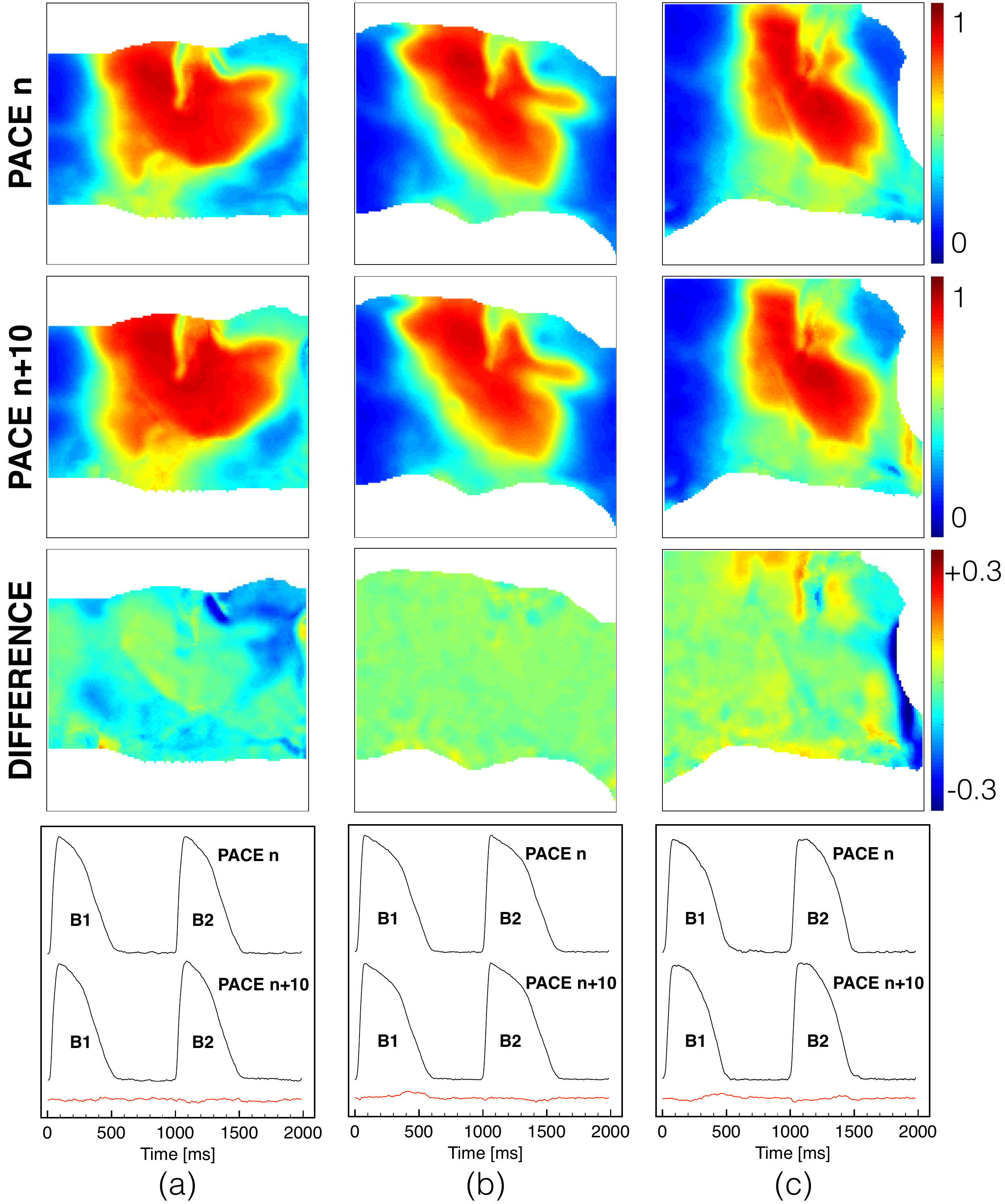}
\end{center}
\caption{{Spatial and temporal comparison of ventricular activation at constant pacing cycle length of $1\,{\rm s}$ under different mechanical loadings (free (a), horizontal (b) and vertical (c) stretch as in Fig.\ref{fig:0}). The first two rows show the spatial distribution of the normalized voltage for beat $n$ and beat $n+10$ with the corresponding difference in the third row (color code is indicated). The last row indicates the time course of a representative pixel in the center of the field of view for two consecutive beats $n$ and $n+10$ with the corresponding difference provided in the red trace.}}
\label{fig:static}
\end{figure}

{Also, in Fig.~\ref{fig:static} we demonstrate that the tissue is at steady-state 
for the selected stimulation rate providing a quantitative comparison of the
spatial and temporal activation sequences.
In particular, after several activations ($>5$), beat $n$ and beat $n+10$ are shown for a selected frame in terms of normalized AP distribution and its spatial difference, as well as comparing the
time course of two consecutive activations (B1, B2) for a representative pixel under the field of view.
In both cases, the spatio-temporal differences recorded are within the physiological variability of a ventricular wedge, the tissue shows a steady-state regime which is considered at resting state for the numerical model.}

Clear MEF effects evidenced in the previous experimental
exercise suggest the incorporation of deformation and stress into the
conduction properties of the cardiac tissue itself.
{The preliminary character of the proposed minimal model implies that 
we do not take into account the intrinsic structural variability of the tissue, but 
we stress that these effects will be investigated in future validation works. Accordingly,}
as a base line model, in the present study we will adapt the formulation recently
proposed in \cite{cherubini:2017} and designed for general purpose
stress-diffusion couplings. Doing so will allow us to readily and
selectively incorporate two main MEF-related mechanisms into
the computational modeling of cardiac electromechanics: 
(i) stretch-activated currents {(SAC)} and (ii) stress-assisted diffusion {(SAD)}. 
The first \emph{paradigm} 
relates the deformed mechanical state to the excitability of the
medium via additional reaction functions (ionic-like currents);
whereas the second one collects the homogenized effects of the
deformation field on the diffusion processes {originating the voltage membrane}.

Within such a framework, we expect stretch-activated currents
and stress-assisted diffusion to counterbalance each other by locally
enhancing tissue excitability as well as smoothing the excitation wave 
according to the mechanical state of the tissue.
In particular, since an external loading activates
SAC at locations where the stretch is high and, at the same time,
induces an heterogeneous and anisotropic diffusion tensor via the SAD mechanisms, 
our study focuses on the role of different mechanical boundary conditions
in affecting action potential propagation and onset of arrhythmias.
Accordingly, these two  MEF mechanisms will be studied numerically in terms of three basic lines.
First, by conducting a parametric analysis of the competing nonlinearities 
such to identify the limits of applicability of the proposed models. 
In particular, we identify in the SAD mechanisms
the most reliable modeling approach
able to reproduce the experienced conduction velocity reduction
upon an applied static loading state.
Then, by performing a selective investigation of spiral onset protocols 
we will characterize the additional nonlinearities that arise
due to MEF. Here we identify 
the different time span of the vulnerable window
obtained via an S1S2 excitation protocol.
Finally, by means of long-run analyses of arrhythmic
scenarios, we compare and contrast static and dynamic
displacement and traction loadings on a two-dimensional,
idealized tissue slab. In this regard, we show how
spiral core meandering results highly affected by the mechanical state
and becomes unstable when SAC and {SAD} parameters are stronger.

Our results highlight several
interesting conclusions regarding the propagation of the excitation
wave in the presence of two {competitive} MEF effects.  
These findings call for novel and
additional experimental investigations. Finally, we provide a thorough
discussion of the applicability of the proposed modeling approach and
its extensions towards more realistic and multiphysics scenarios.

\section{Methods}
The classical stress-assisted formulation proposed in \cite{aifantis:1980} was
developed in the context of dilute solutes in a solid. 
A similarity exists 
between this fundamental process and the propagation of voltage membrane within cardiac tissue. 
Indeed,  on a macroscopically
rigid matrix, the propagating membrane voltage can be regarded as a   
continuum field undergoing slow diffusion.  Here we consider a similar approach (developed 
in \citealp{cherubini:2017}) which generalizes Fick's diffusion by using
the classical Euler's axioms of continuously distributed matter. In
particular, the balance of momentum can be imposed such to ensure
frame invariance, a property of high importance in mechanical
applications~\citep{tadmor:2012}. We also assume quasi-static
conditions for the continuum body, such that its macroscopic
response is, in principle, independent from the diffusion process. On
the contrary, the diffusion process will strongly depend on the
mechanical state of the tissue.

\subsection{Continuum electromechanical model}
We will assume that the body is a hyperelastic material and its
motion will be described using finite kinematics. We will adopt an
indicial notation where repeated indices indicate summation. We
identify the relationship between material (reference), $X_I$, and
spatial (deformed), $x_i$, coordinates via the smooth map
$x_i(X_I)$. The deformation gradient tensor $F_{iI}=\partial x_i
/ \partial{X_I}$ allows to determine further properties of the
continuum's motion. We indicate with $J=\det F_{iI}$ the Jacobian of
the map and with $C_{IJ}=F_{kI}F_{kJ}$ and $B_{ij}=F_{iK}F_{jK}$ the
right and left Cauchy-Green deformation tensors, respectively. We
assume that the generic myocardial fiber direction (the unit vector
characterizing the microstructural property of the continuum body) in
the material configuration, $a_I$,
is mapped to the deformed configuration as $a_i=F_{iJ}a_J$ such that
we can define the current fiber $a_i = a_I/\lambda$. Following the
standard frame indifference mechanical framework~\citep{spencer:1980},
these quantities are related to the invariants of the deformation
in the following manner
\begin{equation}
	I_1 = C_{II} \,, \quad
	I_2 = \dfrac{1}{2}\left[ (C_{II})^2 - C_{IJ} C_{JI} \right] \,, \quad
	I_3 = \det C_{IJ}=J^2 \,, \quad
	I_4 = C_{IJ}a_I a_J=\lambda^2 \,. 
\end{equation}
The principal invariants $I_1$ and $I_2$ rule the deviatoric response
of the medium, the third invariant $I_3$ quantifies volumetric changes
of the material, while the fourth pseudo-invariant $I_4$ measures the
directional \emph{fiber stretch}, $\lambda$. This last entity is
intrinsically directional, so for two-dimensional models, we will
simply assign a horizontal myocardial direction
$(1,0)^T$. In what follows, the symbol $\delta_{ij}$
denotes the second-order identity tensor.

As anticipated above, we will base our model on the
stress-assisted diffusion formulation from \cite{cherubini:2017}. We
do however, generalize the governing equations adopting a
more accurate nondimensional three-variable model of cardiac action
potential (AP) propagation introduced in \cite{fenton:1998}, and we
will account for SAC~\citep{panfilov:2005},
that were not considered in~\cite{cherubini:2017}. Even though
several more physiological assumptions could be made, here we will
focus on a purely phenomenological approach~\citep{fenton:2008}.

In the deformed configuration, the chosen electrophysiological
model consists of three variables: the membrane potential $u$, and a
fast and slow transmembrane ionic gates $v,w$. They satisfy
the following RD system
\begin{subequations}
	\label{eq:FK}
\begin{align}
	\label{eqFKu}
	\frac{\partial u}{\partial t} &=
	\frac{\partial}{\partial x_i}\left( d_{ij}(\sigma_{ij}) \frac{\partial u}{\partial x_j}\right)
	- {\rm I}_{\rm ion} (u,v,w)
	+ \Isa
	+ {\rm I}_{\rm ext} \,,
	\\
	\label{eqFKv}
	\frac{d v}{d t} &=
	(1-H_c) \left(\frac{1-v}{\tau_v^-}\right)- H_c \frac{v}{\tau_v^+} \,,
	\\
	\label{eqFKw}
	\frac{d w}{d t} &=
	(1-H_c) \left(\frac{1-w}{\tau_w^-}\right)- H_c \frac{w}{\tau_w^+} \,,
\end{align}
\end{subequations}
where Neumann zero-flux boundary conditions are imposed for \eqref{eqFKu},  
i.e.  $[d_{ij}{\partial u}/{\partial x_j}] n_i = 0$, where $n_i$ is
the outward normal on the domain boundary.  System~\eqref{eq:FK} describes the
propagation of a normalized dimensionless membrane potential, which
can be mapped to physical quantities as $u=\left(V_m -
V_o\right)/\left(V_{fi} - V_o\right)$ (see\,\citealp{fenton:1998} for
details) where $V_m$ stands for the physical transmembrane potential,
$V_o$ is the resting membrane potential and $V_{fi}$ represents the
Nernst potential of the fast inward current.  In Eq.~\eqref{eqFKu},
the total transmembrane density current, ${\rm I}_{\rm ion} (u,v,w)$,
is the sum of a fast inward depolarizing current, ${\rm I}_{fi}
(u,v)$, a slow time-independent rectifying outward current, ${\rm
I}_{so} (u)$, and a slow inward current, ${\rm I}_{si}
(u,w)$, given by
\begin{align*}
  {\rm I}_{fi} (u,v) &= -\frac{v}{\tau_d} H_c\left(1-u\right)\left(u-u_c\right)\,, \\
  {\rm I}_{so} (u) &= \frac{u}{\tau_o}\left(1- H_c\right)+\frac{1}{\tau_r} H_c\,, \\
  {\rm I}_{si} (u,w) &= -\frac{w}{2\tau_{si}}\left(1+\tanh\left[k\left(u-u_c^{si}\right)\right]\right)\,,
\end{align*}
where $\tau_v^-\left(u\right)=H_v\tau_{v1}^-+\left(1-H_v\right)\tau_{v2}^-$
is the time constant governing the reactivation of the fast inward
current, and $H_x=H_x\left(u-u_x\right)$ is the standard Heaviside
step function.  
${\rm I}_{\rm ext}$ is the space and time-dependent external stimulation current with amplitude ${\rm I}_{\rm ext}^{\max}$.
All model parameters are collected in Table~\ref{tabTpar}.

\begin{table}[t!]
\bigskip{}\centering{}
\begin{tabular}{rlrlrlrl | ll}
\hline
\hline
$\bar{g}_{fi}$& 4  & $\tau_d$ &  $C_m/\bar{g}_{fi}$ &$\tau_w^+$   & 667&$\epsilon_0$ & 0.1& $u^{\rm init} = 0$\\
$\tau_r$     & 50 &$C_m$ & 1 $\mu{\rm F/cm}^2$ &$\tau_w^-$   & 11 &$k_{Ta}$ & 9.58&$v^{\rm init} = 1$\\
$\tau_{si}$   & 45 &$V_o$ & $-85$&$u_c$ & 0.13 &$c_1$ & 6&$w^{\rm init} = 1$\\
$\tau_o$     & 8.3&$V_{fi}$ & $15$ &$u_v$ & 0.055 &$c_2$ &  2&${\pmb{\varphi}}^{\text{init}} = 0$\\
$\tau_v^+$   &3.33&$D_0$ & $1 \cdot 10^{-3}$ &$u_c^{si}$ & 0.85 &$G_s$ & $[0 ; 0.25]$&$p^{\text{init}} = 0$\\
$\tau_{v1}^-$ &1000&$D_1$ & $[-1.5; 0] \cdot10^{-4}$ &$k$ & 10 &$\usa$ & 0.4&$T_a^{\text{init}} = 0.2$\\
$\tau_{v2}^-$ &19.6&$D_2$ & $1 \cdot 10^{-5}$ & $\rm I_{ext}^{\max}$ & 2 & $t_{\max}$ & 9 &\\[1.1ex]
\hline
\hline
\end{tabular}
\caption{\label{tabTpar}
Model parameters for the electromechanical three-variable model,
considered as in~\citep{fenton:1998,cherubini:2017}.  Time units are
$\rm ms$, length is $\rm cm$, the term $\bar{g}_{fi}$ is in ${\rm
mS/cm}^2$, dimensional voltages are in $\rm mV$, and stiffness in
$\rm MPa$. Square brackets indicate range of parameter
variability, and the rightmost column specifies initial conditions for
a resting tissue.}
\end{table}

The mechanical problem, stated also on the current
configuration and occupying the domain $\Omega(t)$, respects the
balance of linear momentum and mass, written in terms of displacement, 
$\pmb{\varphi}$, and pressure, $p$, and set in a quasi-static form.  The
problem is complemented with displacement and traction boundary
conditions set on two different parts of the boundary $\Gamma_D$ or
$\Gamma_N$:
\begin{subequations}
\label{eq:Equil}
\begin{alignat}{2}
    \dfrac{\partial \sigma_{ij}}{\partial x_i} = 0 
\quad \text{and} \quad \rho d\hat{v} & = \rho_0d\hat{V}, \quad 
&\text{in}\quad \Omega(t),\label{momentum-mass}   \\
    \pmb{\varphi} & = \tilde{\pmb{\varphi}}(t) , \quad 
& \text{on}\quad \Gamma_D(t), \label{bc-displ} \\
\sigma_{ik} n_k & = \tilde{t}_i (t) , \quad 
&\text{on}\quad \Gamma_N(t), \label{bc-traction}
  \end{alignat}
\end{subequations}
where $\rho_0,\rho$ and
$d\hat{v},d\hat{V}$ are the densities and volumes of the solid in the
undeformed and deformed configurations, respectively.  In \eqref{bc-displ}, 
$\tilde{\pmb{\varphi}}(t)$ is a known (possibly time-dependent)
displacement and in \eqref{bc-traction}, $\tilde{t}_i (t)$ is a possibly time-dependent
traction force. In both cases, the tissue is stretched up to a
maximum level of 20\% of the resting length such to activate all MEF
components. In addition, the time-variation of the imposed boundary
conditions is much slower than the governing dynamic physical
processes, and therefore a quasi-static mechanical equilibrium is
maintained.

The two sub-problems \eqref{eq:FK},\eqref{eq:Equil} are completed via
the following mixed constitutive prescriptions for incompressible
isotropic hyperelastic materials $(J=1)$:
\begin{subequations}
\begin{alignat}{3}
    \sigma_{ij} &= 
    2 c_1 B_{ij} - 2 c_2 B^{-1}_{ij} - p \delta_{ij} + 
    T_a \delta_{ij} \,,
    \label{eq:Stress}
    \\
    \dfrac{\partial T_a}{\partial t} &= 
    \epsilon(u) (k_{T_a} u - T_a) \,,
    \label{eq:AStress}
    \\
    d_{ij}(\sigma_{ij}) &= 
    D_0 \delta_{ij} + D_1 \sigma_{ij} + D_2 \sigma_{ik}\sigma_{kj} \,,
    \label{eq:Diffusivity}
    \\
    \Isa &= G_s H_{\rm sac} (\lambda-1) (\usa - u) \,.
    \label{eq:SAC}
\end{alignat}
\end{subequations}
Equation~\eqref{eq:Stress} specifies a 
constitutive form for the Cauchy stress tensor 
(total equilibrium stress in the current deformed configuration) 
highlighting two multiscale contributions on the tissue 
deformation. First, the passive material response follows that of an
incompressible Mooney-Rivlin hyperelastic solid and it is 
characterized by two stiffness parameters $c_1$ and $c_2$; and 
secondly, the active component contributing to the total stress 
in the form of an additional hydrostatic force with amplitude $T_a$. 
The dynamics of $T_a$ are described by Eq.~\eqref{eq:AStress}, 
where the constant $k_{Ta}$ modulates the amplitude of the 
active stress contribution, while $\epsilon(u)$ is a contraction switch function: 
$\epsilon(u) = \epsilon_0$ if $u<0.005$, and 
$\epsilon(u) = 10\epsilon_0$ if $u\ge0.005$.

Equation~\eqref{eq:Diffusivity} characterizes the stress-assisted
diffusion contribution describing the effect of tissue deformation on
the AP spreading. The parameter $D_0$ represents the usual diffusion
coefficient for isotropic media, i.e. diffusivity = [L$^2$ T$^{-1}$], while $D_1$ and $D_2$ introduce the impact of
mechanical stress through linear and nonlinear contributions,
respectively, on the diffusive flux. Accordingly, $D_1$ and $D_2$ have
units of [L$^2$ T$^{-1}$ P$^{-1}$] and [L$^2$ T$^{-1}$ P$^{-2}$], respectively. We
also remark that Eq.~\eqref{eq:Diffusivity} reduces to the classical
diffusion equation for $D_1\equiv D_2=0$.

Finally, Eq.~\eqref{eq:SAC} describes the stretch-activated current
contribution (which is usually adopted as the sole MEF effect). The
term $\Isa$ affects the ionic (reaction) currents in the
electrophysiological system and is formulated as a linear function of
the membrane potential $u$ and the fiber stretch $\lambda$. Here,
$G_s$ modulates the amplitude of the current, $\usa$ represents a
referential (resting) potential while, $H_{\rm sac}$ is a switch
activating this additional reaction current only when the myocardial
fiber is elongated, i.e. $H_{\rm sac}=1$ for $\lambda \ge 1$ and
$H_{\rm sac}=0$ for $\lambda<1$.

We also introduce the definition of spiral tip
(core of the spiral wave) as the point with instantaneous null velocity
(see \citep{fenton:1998} for details). In practice, for two-dimensional
domains, we choose an isopotential line of constant membrane
voltage, $u(R_I,t)=u_{\rm iso}$, where 
$R_I=x_{\rm tip} X_I + y_{\rm tip} Y_I$ represents the 
position vector in the reference undeformed configuration
identifying the boundary between depolarized and repolarized 
regions. Accordingly, the spiral tip can be defined as the 
point in space where the excitation front meets the 
repolarization waveback of the action potential,
conforming with the operative definition:
\begin{align}
    u(R_I,t) - u_{\rm iso} = \dfrac{\partial u(R_I,t)}{\partial t} \equiv 0 \,. 
\end{align}
We numerically identify the tip coordinates $(x_{\rm tip}, y_{\rm tip})$
by considering $u_{\rm iso}=0.5$ with tolerance of $10^{-4}$.

\subsection{Numerical approximation}\label{sec:numerics}
The electromechanical problem is written in the undeformed
configuration and subsequently computationally solved via a finite
element method. {Even if the model originates as an extension of our 
contribution in \cite{cherubini:2017}, the numerical method employed 
here is simpler, as we do not solve for stresses explicitly but rather postprocess 
them from the computed discrete displacements. The overall numerical scheme 
for active stress electromechanics with SAC is therefore not 
precisely novel, but will still provide a few details for sake of 
completeness of the presentation and future reproducibility of results. 
Further details could be found in e.g.~\cite{ruiz:2015}. 
We discretize displacements with vectorial 
piecewise quadratic and continuous polynomials, and the
pressure field using Lagrangian finite elements (that is, the classical Taylor-Hood 
method). All remaining
unknowns (associated to the electrophysiology and to the active
tension) are also approximated using piecewise linear and continuous
elements.  
Let us then consider a regular, quasi-uniform partitions $\cT_h$ 
 of $\overline{\Omega(0)}$ into triangles $T$ of 
diameter $h_T$, where $h = \max\{ h_T:\, T\in \cT_h\}$ is 
the meshsize. The finite element spaces mentioned above 
are defined as (see e.g. \citealp{qv94}) 
\begin{align*}
\bH_h & := \{\bpsi \in \bH^1(\Omega(0)) : \bpsi|_T \in [\mathbb{P}_{2}(T)]^2\ 
\forall T\in\cT_h, \text{ and } \bpsi = \boldsymbol{0} \text{ on } \Gamma_D(0)\},\\
Q_h &: = \{ q\in L^2(\Omega(0)) \cap C^0(\Omega(0)): q|_T \in\mathbb{P}_1(T)\   
\forall T\in\cT_h\},\\
W_h &: = \{ \psi \in H^1(\Omega(0)): \psi|_T \in\mathbb{P}_{1}(T)\   
\forall T\in\cT_h\},
\end{align*}
for the case of clamped boundaries at $\Gamma_D(0)$.} 

Let us also construct an equispaced partition of the 
time domain $0=t^0 < t^1 = \Delta t < \cdots < t^M = t_{\max}$. 
The coupled problem is solved sequentially between the
mechanical and electrochemical blocks. 
A description of the needed computations at each time 
step $t^n$ is as follows:

\noindent{\textbf{\underline{Step 1:}} From the known values $u_h^n,v_h^n,w_h^n, T_{a,h}^n, D_h^n, \lambda_h^n$, find 
$u_h^{n+1},v_h^{n+1},w_h^{n+1}, T_{a,h}^{n+1}$ such that} 
\begin{align*}
\int_{\Omega(0)} \frac{u_h^{n+1}}{\Delta t}   \psi^u_h 
+   \int_{\Omega(0)} D_h^n \nabla u_h^{n+1} \cdot \nabla \psi^u_h & = 
\int_{\Omega(0)}  [\frac{u_h^{n}}{\Delta t}  + \mathrm{I_{ion}}(u_h^{n}, v_h^n,w_h^n) 
+ \mathrm{I_{sac}}(\lambda_h^n,u_h^n) +\mathrm{I_{ext}}]  \psi^u_h  ,\\
\frac{1}{\Delta t} \int_{\Omega(0)}  v_h^{n+1} \psi^v_h & = \int_{\Omega(0)}  [\frac{1}{\Delta t} v_h^{n} +f_v(u_h^{n}, v_h^n) ] \psi^v_h ,\\
\frac{1}{\Delta t} \int_{\Omega(0)}  w_h^{n+1} \psi^w_h & = \int_{\Omega(0)}  [\frac{1}{\Delta t} w_h^{n} +f_w(u_h^{n}, w_h^n) ] \psi^w_h,\\
\frac{1}{\Delta t} \int_{\Omega(0)}  T_{a,h}^{n+1} \psi^{T_a}_h & = 
\int_{\Omega(0)}  [\frac{1}{\Delta t} T_{a,h}^{n} +f_{T_a}(u_h^{n}, T_{a,h}^n) ] \psi^{T_a}_h,
\end{align*}
{for all  $(\psi^u_h,\psi^v_h,\psi^w_h,\psi^{T_a}_h) \in [V_h]^4$. This scheme for the 
electric/activation system 
 is given in a first-order semi-implicit form: the nonlinear 
reaction terms and the coupling stress-assisted diffusion are taken explicitly, while the linear part of 
diffusion is advanced implicitly. Here}  
\[ {D_h^n = D_0 \bC^{-1}(\bphi_h^n) + \frac{D_1}{J(\bphi_h^n)} \bS(\bphi_h^n) + \frac{D_1}{J(\bphi_h^n)^2} \bS(\bphi_h^n)^2, \quad 
\lambda_h^n = \sqrt{C_{11}(\bphi_h^n)},} 
\]
{are the explicit approximation of the stress-assisted diffusivity and of the 
stretch in the fiber direction, all in the reference configuration.} 

\noindent{\textbf{\underline{Step 2:}} Given the 
the activation value $T_{a,h}^{n+1}$ computed in Step 1 of this iteration, solve the 
nonlinear elasticity equations} 
\begin{align*}
\int_{\Omega(0)}  \bF(\bphi_h^{n+1}) \bS(\bphi_h^{n+1},p_h^{n+1},T_{a,h}^{n+1}) : 
\nabla \bpsi_h  & = 0 & \quad \forall  \bpsi_h \in \bH_h,\\
  \int_{\Omega(0)} q_h [J(\bphi_h^{n+1}) -1]   & = 0 & \quad \forall  q_h \in Q_h,
\end{align*}
{where} 
\[\bS = 2[c_1 + c_2 \tr(\bC(\bphi_h^{n+1}))] \bI - 2c_2\bC(\bphi_h^{n+1}) - 
p_h^{n+1}J(\bphi_h^{n+1}) \bC^{-1}(\bphi_h^{n+1})  + T_{a,h}^{n+1} \bC^{-1}(\bphi_h^{n+1}) ,\]
{is the second Piola-Kirchhoff stress tensor.} 

\noindent{\textbf{\underline{Step 3:}} The solution of the problem in Step 2 uses a Newton-Raphson method 
whose iterations are terminated once the energy residual drops below the relative
tolerance of 1$\cdot10^{-6}$. The solution to each linear tangent
problem is conducted with
the BiCGSTAB method preconditioned with an incomplete LU
factorization. The iterations of the Krylov solver are terminated
after reaching the absolute tolerance 1$\cdot10^{-5}$.  The residual
computation for the mechanical problem also contains the terms arising
from time-dependent displacement or traction boundary conditions,
which also need to be assigned at each timestep. For instance, in an
uniaxial test (denoted dynamic displacement in the examples below), 
the left segment of the boundary is clamped (zero 
displacements are imposed), the bottom and top edges are subject to
zero normal stress, and the right edge is pulled according to the
displacement $\tilde{\pmb{\varphi}}(t) = \left[0.2L \sin^2(\pi/400 \, t),0 \right]^T$.}

\begin{figure}[t!]
\begin{center}
	\includegraphics[width=0.65\textwidth]{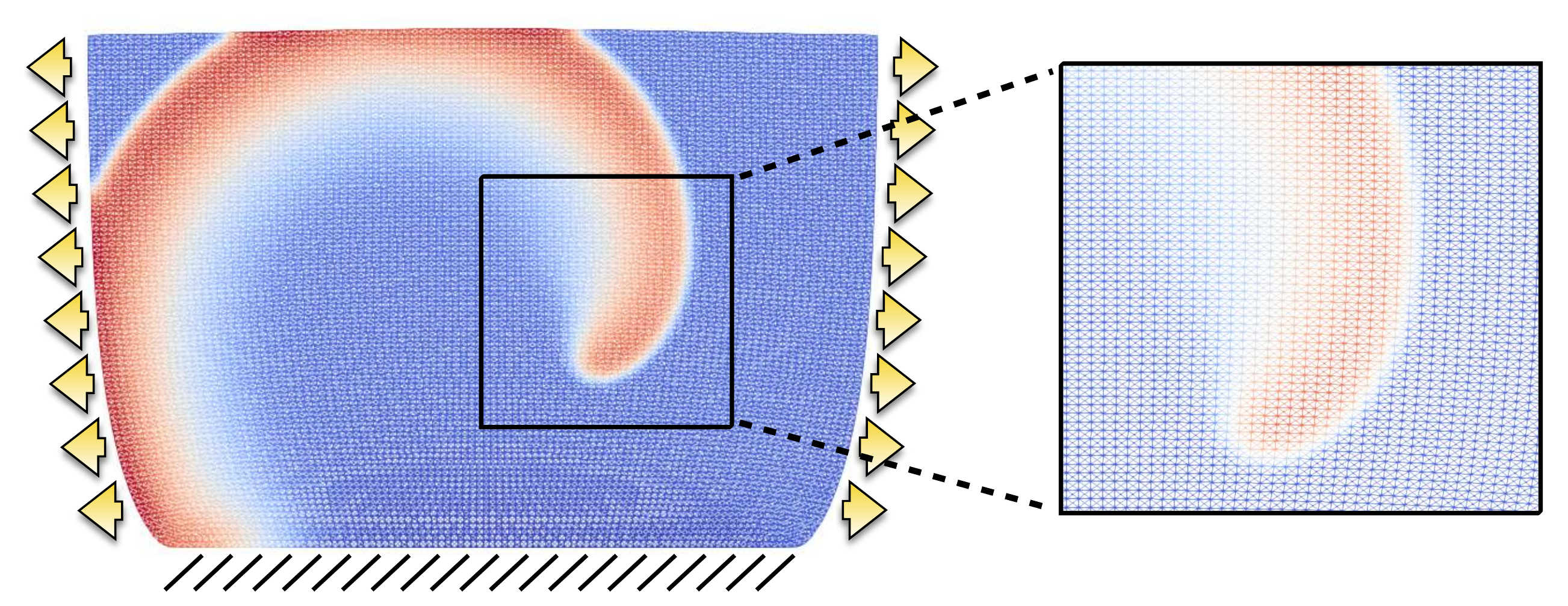}
\end{center}
\caption{Example of structured mesh employed in the computational
results. The skeleton grid is displayed on the deformed configuration
when the domain is subject to traction (arrows) and fixed displacement (lines) boundary
conditions, and a zoom exemplifies the mesh resolution for a rather
coarse spiral front.}
\label{fig:mesh}
\end{figure}

All tests are conducted using a two-dimensional
slab of dimensions $L\times L=6.2\times6.2\,{\rm cm}^2$, which is the
same configuration used to produce the dynamics analyzed in
\cite{fenton:1998}. 
The computational domain is discretized with a structured
triangular mesh of 10000 elements. After a mesh convergence test
involving conduction velocities and reproducing the expected values
for planar excitation waves reported in \cite{fenton:1998}, we
proceeded to fix the temporal and spatial resolutions to
$\Delta t=0.1\,{\rm ms}$, $h=0.062\,{\rm cm}$, respectively.  A
representative example of the mesh is provided in Fig.~\ref{fig:mesh},
plotted in the deformed configuration under both traction and
displacement boundary conditions and highlighting the spiral wave
resolution.  All numerical tests were carried out using the
open-source finite element library FEniCS~\citep{fenics}.

\section{Results}\label{sec:results}
In the following, we adopt a parametric setup fitted for the modified
Beeler-Reuter model \eqref{eq:FK}, while selectively changing MEF
parameters $(D_1,G_s)$. This choice provides a 
{reference, unloaded, model configuration with 
constant CV of 0.42 ${\rm m/s}$} and a circular meandering for a 
free spiral on a homogeneous and isotropic domain. 
{Such values deviate as the MEF coupling is activated.}

\subsection{Conduction velocity analysis}
We start analyzing the parameter space associated to the two
MEF contributions in our model. That is, the stress-assisted
coefficients $D_1,D_2$ and the SAC amplitude
$G_s$. The study will be restricted to a static homogeneous
stretched state (e.g. a uniaxial Dirichlet boundary condition
$\boldsymbol{\varphi}=(0.2 L,0)^T$ set on the right edge of the
domain). All remaining material and electrophysiology parameters will
be kept constant, except that we fix the relative influence of
the nonlinear contribution in the stress-assisted diffusion, by
setting $D_2$ to be one order of magnitude smaller than $D_1$. This
configuration will highlight MEF effects in a minimal, but still
comprehensive manner.

\begin{figure}[h!]
\begin{center}
	\includegraphics[width=0.99\textwidth]{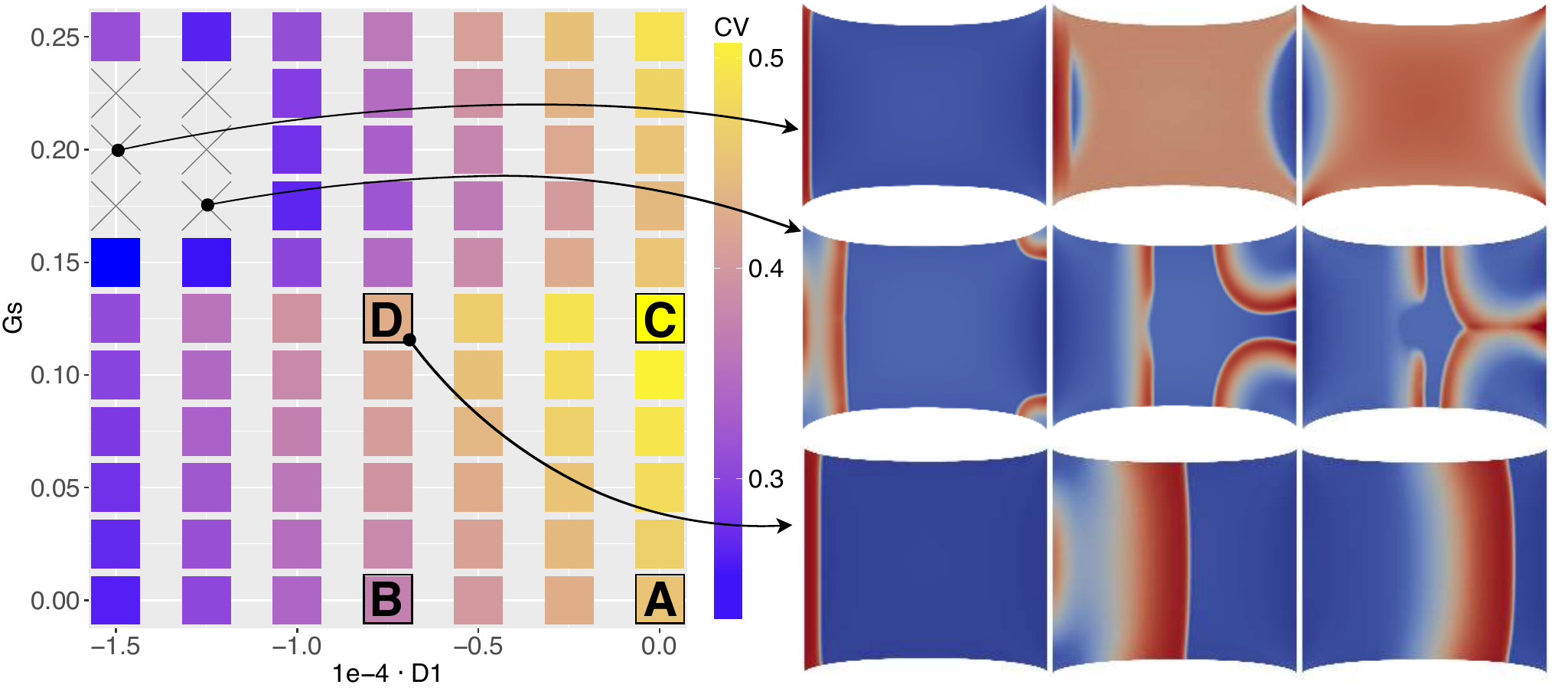}
\end{center}
\caption{MEF parameter space associated to the conduction velocity measured 
on the propagating front of a planar excitation wave (stimulation on the left edge and propagation towards the right boundary) elicited on a static 
uniaxially stretched domain (CV in $[\rm m/s]$). 
Four selected combinations of MEF parameters (A,B,C,D, in Tab.~\ref{tab:s1_s2})
are highlighted together with two additional cases in which CV was not recorded.
{On the right, three consecutive time frames of the activation are selected.}}
\label{fig:param}
\end{figure}

\begin{figure}[htp!]
\begin{center}
	\includegraphics[width=0.95\textwidth]{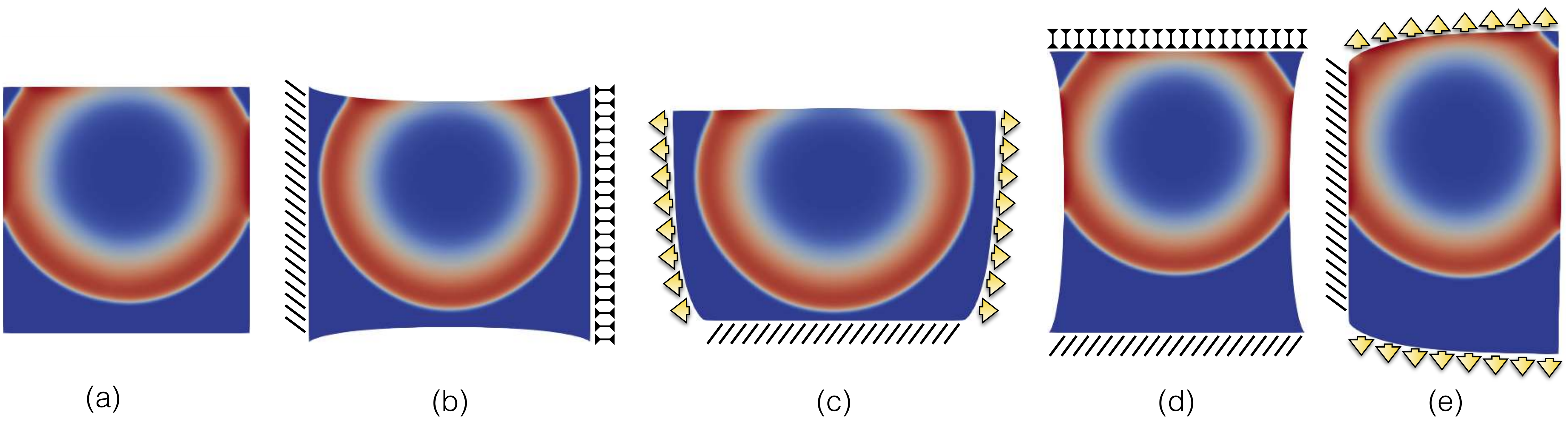}
\end{center}
\caption{Point-wise activation frame for five different static boundary conditions qualitatively
reproducing ventricle wedge preparation measurements considering the parameter combination
$(D_1,G_s)=(-0.75\cdot 10^{-4},0)$:
(a) free edges,
(b) horizontal displacement,
(c) vertical displacement,
(d) horizontal traction,
(e) vertical traction.
Color code refers to the normalized action potential.
}
\label{fig:point}
\end{figure}

Fig.~\ref{fig:param} portrays the conduction velocity obtained
for all combinations of $(D_1,G_s)$ on the parameter space. The
quantity is measured as the wave-front velocity of a planar
excitation wave along its propagation. The plot illustrates the variability of the
recorded CV amplitude (in the range 0.25 -- 0.5 ${\rm m/s}$)
according to the MEF coupling intensity variation 
and to histogram measures in Fig.~\ref{fig:stat}. 
In particular, starting from a physiological baseline of 0.42 ${\rm
m/s}$, when neither {SAC} nor  {SAD}
is present ($D_1=0,G_s=0$), we observe a net increase of CV
for $(D_1=0, G_s>0)$ while we recover CV decrements for $(D_1<0,
G_s=0)$. This specific aspect reproduces what is expected from
experimental evidence, i.e., MEF decreases the CV of the excitation
wave~\citep{ravelli:2003}.
\\
Besides, for higher values of $G_s$, we
obtain two unexpected results.  First, for $G_s>0.15$ we observe a
decrement of CV for different values of $D_1$.  Second, for the
particular combination $(D_1<-10^{-4}, G_s>0.15)$ the wave disappears
from the domain or annihilates due to excessive activation (see
e.g. side panels in Fig.~\ref{fig:param} or the top row in Fig.~\ref{fig:cases}).
Consequently, we are not able to measure any propagation 
(which reflects in the {combinations with $\times$} of the figure).
This last result is somehow counterintuitive 
since, as evidenced by Fig.~\ref{fig:0}, we experimentally experience a complete 
depolarization of the tissue with AP propagation, in the case of 
fixed stretch.
To support this point, in Fig.~\ref{fig:point} we provide a representative sequence of
point-wise activations delivered on our simplified 2D domain and mimicking
the experimental protocol conducted in Fig.~\ref{fig:0} for a selected parameter choice,
i.e. $(D_1,G_s)=(-0.75\cdot 10^{-4},0)$.
In this case, the AP excitation wave propagates differently
according to the applied stretch state, both horizontal and vertical displacement and traction.
In addition, the computed CVs change similarly to what observed in Fig.~\ref{fig:stat}.
We remark that such a comparison with experimental observations
is purely qualitative and does not represent a validation of the model.

\subsection{S1-S2 excitation protocol}
We further investigate the strength of MEF coupling effects. In
particular, we want to determine which specific contribution
(stretch-activated currents or stress-assisted diffusion) exhibits a
better match against experimental evidence, and for this we assess
changes in the S1-S2 stimulation protocol. In practice, in
order to induce a spiral wave on an excitable tissue, one
typically generates a planar electrical excitation (S1), followed by
a second broken stimulus (S2) during the repolarization phase of the
S1 wave, the so called vulnerable window~\citep{karma:2013}. 
In our case, we selected a reduced set of MEF
parameters $(D_1,G_s)$ indicated in Tab.~\ref{tab:s1_s2} as
A,B,C,D.
These values are motivated by the results from
Fig.~\ref{fig:param}.  In particular, we select only the parameter
combinations that produce either a unique decrement or
increment of CV.

\begin{table}[ht!]
\centering
\begin{tabular}{lllll}
\hline
\hline
& $D_1$ & $G_s$ & {CV $[\rm{m/s}]$} & $t_{\rm S_2}^{\rm min} - t_{\rm S_2}^{\rm max}$ $[{\rm ms}]$ $\vphantom{\in^X_X}$\\[0.5ex]
\hline 
A: & 0 & 0 & {0.45} & 225 - 240 $\vphantom{\in^X}$\\
B: & $-0.75\cdot 10^{-4}$ & 0 & {0.36} & 243 - 255 \\ 
C: & 0 & 0.125 & {0.42} & 133 - 147 \\ 
D: & $-0.75\cdot 10^{-4}$ & 0.125 & {0.52} & 143 - 157 \\ 
\hline
\hline
\end{tabular}
\caption{\label{tab:s1_s2} 
Parameter calibration associated to the S1-S2 protocol. 
Combination of MEF parameters $(D_1,G_s)$, 
{corresponding CV},
minimum, $t_{\rm S_2}^{\rm min}$, and maximum, $t_{\rm S_2}^{\rm max}$, stimulation time
required for spiral wave onset {(vulnerable window)}.}
\end{table}

\begin{figure}[t!]
\centering
\includegraphics[width=0.9\textwidth]{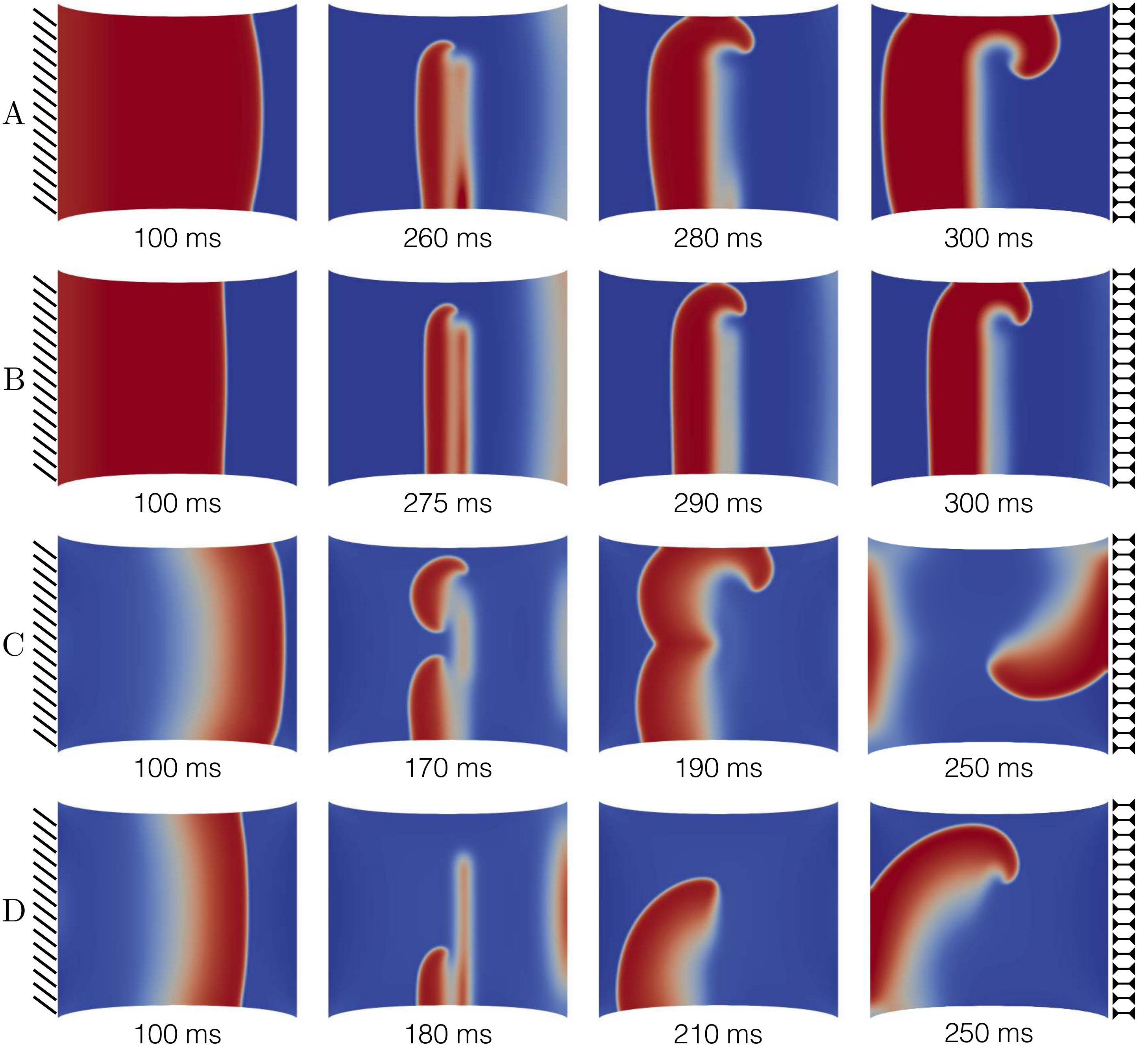}
\caption{S1-S2 stimulation protocol applied on a static uniaxial
stretched configuration for different combinations of MEF parameters
$(D_1,G_s)$ as provided in Tab.~\ref{tab:s1_s2}. The color code refers to
normalized dimensionless membrane potential, $u$, (blue-red mapped to
[0-1]). {Selected time frames are provided in the subpanels.}}
\label{fig:S1S2}
\end{figure}

Figure \ref{fig:S1S2} shows the different dynamics obtained via the
S1-S2 protocol for the four different sets of MEF parameters. The
first column is set at $100\,{\rm ms}$ from the S1 stimulus for all
the combinations, while the remaining frames are selected to highlight
the elicited behavior. As a result, we observe that the deformation
state of the tissue influences the overall dynamics differently.
The first column highlights the variability in the AP
wavelength, {representing the spatial extension of the activation wave}, 
which is due to the different repolarization states of the
tissue induced by stress-assisted diffusion and stretch-activated
currents. {In particular, the AP wavelength varies as
$>6.2\,{\rm cm}$ for case A,
$=6.2\,{\rm cm}$ for case B, and
$<2\,{\rm cm}$ for cases C, D.}
In fact, when the second contribution $G_s$ is present, the
excitation wave is much reduced with respect to the profiles generated
with the electrophysiological three-variable model \eqref{eq:FK} and 
fine-tuned on experimental data. Such an effect is not present
when $G_s=0$.
\\
Secondly, cases A and B (that is, where only $D_1$ is
activated) provide the expected reduction in CV and a similar behavior
for spiral onset. Contrariwise, cases C and D (where also the
contribution of $G_s$ is present) induce much more complex
dynamics, not expected in an isotropic medium. In particular, case C
leads to a wave break and multiple spiral generation at the S2
stimulus that eventually collide and result in a single spiral
wave. On the other hand, case D shows a more stable behavior generated
by the presence of $D_1$.
\\
In addition, Tab.~\ref{tab:s1_s2} also provides the 
{minimum and maximum} delay for
the S2 stimulation {(vulnerable window)}
allowing to induce a spiral wave 
in the uniaxially
stretched tissue. It is evident that the presence of {SAC} 
reduces the {minimum S2 stimulation time, 
$t_{\rm S_2}^{\rm min}$,} by 
about $\rm 100\,ms$
with respect to the other cases
{and slightly increase the overall time span of the vulnerable window}. 
Such a variation is
motivated on the additional reaction current induced by the presence
of $\Isa$ everywhere in the medium, 
but it is not expected from the {experimental} isochrones
{provided} in Fig.~\ref{fig:0}.

\begin{figure}[t!]
\begin{center}
	{\includegraphics[width=0.9\textwidth]{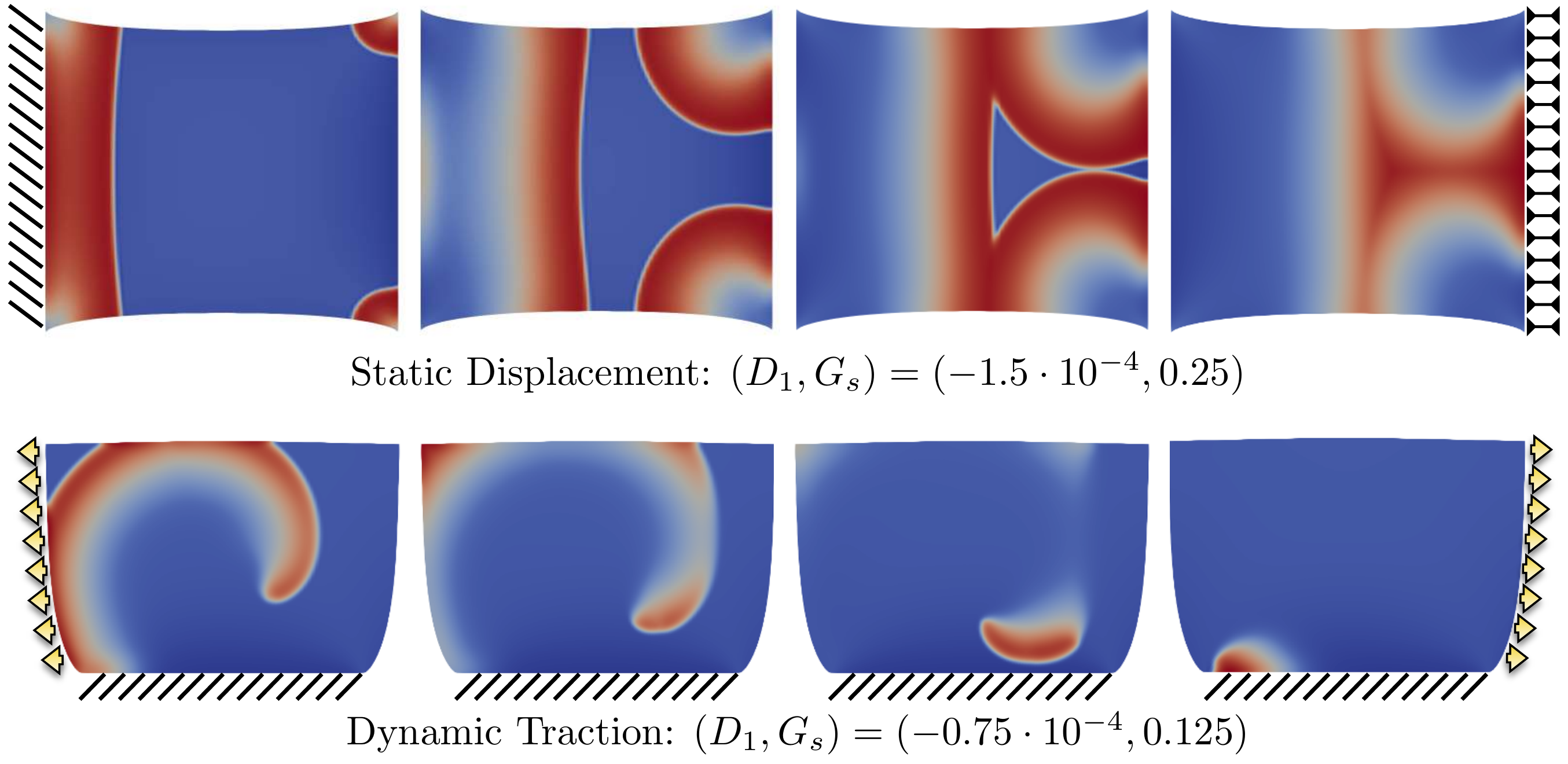}}
\end{center}
\caption{Example of different propagation patterns according to  
different mechanical boundary conditions and parameter space.
First row shows the uniaxial static displacement configuration 
for which the selected parameters induce additional activations
from the corners of the domain due to the excessive level of {SAC} ($G_s$).
Second row shows the dynamic traction configuration
for which the initiated spiral wave goes through breakup due to the effect of
mechanical loading.
}
\label{fig:cases}
\end{figure}

To further corroborate this analysis, we provide in the top panels of 
Fig.~\ref{fig:cases} an additional sequence
referring to the combination $(D_1,G_s)=(-1.5\cdot10^{-4},0.25)$ 
in the case with static displacement boundary conditions, 
which falls in the range where no CV wave was measured. 
As anticipated, an excessive contribution due to {SAC} elicits extra
activations where the stretch is maximum, i.e. at the corners of the
domain. This particular behavior is not obtained when the
stress-assisted contribution $D_1$ is very high.
Next, the bottom panels of Fig.~\ref{fig:cases} show results 
using the combination $(D_1,G_s)=(-0.75\cdot10^{-4},0.125)$, which allows 
the quantification of CV but can eventually lead to spiral breakup 
and non-sustainability of the arrhythmic patterns due to the 
mechanical state of the tissue (corresponding to the case of dynamic traction, 
described below). This is a representative example of the key importance of boundary conditions and how MEF effects 
could be effectively translated into clinical studies.

\subsection{Spiral drift and effects due to boundary conditions}
\label{sec:spiral}
\begin{figure}[h!]
\centering
\subfloat[Static Displacement]{\includegraphics[width=0.48\textwidth]{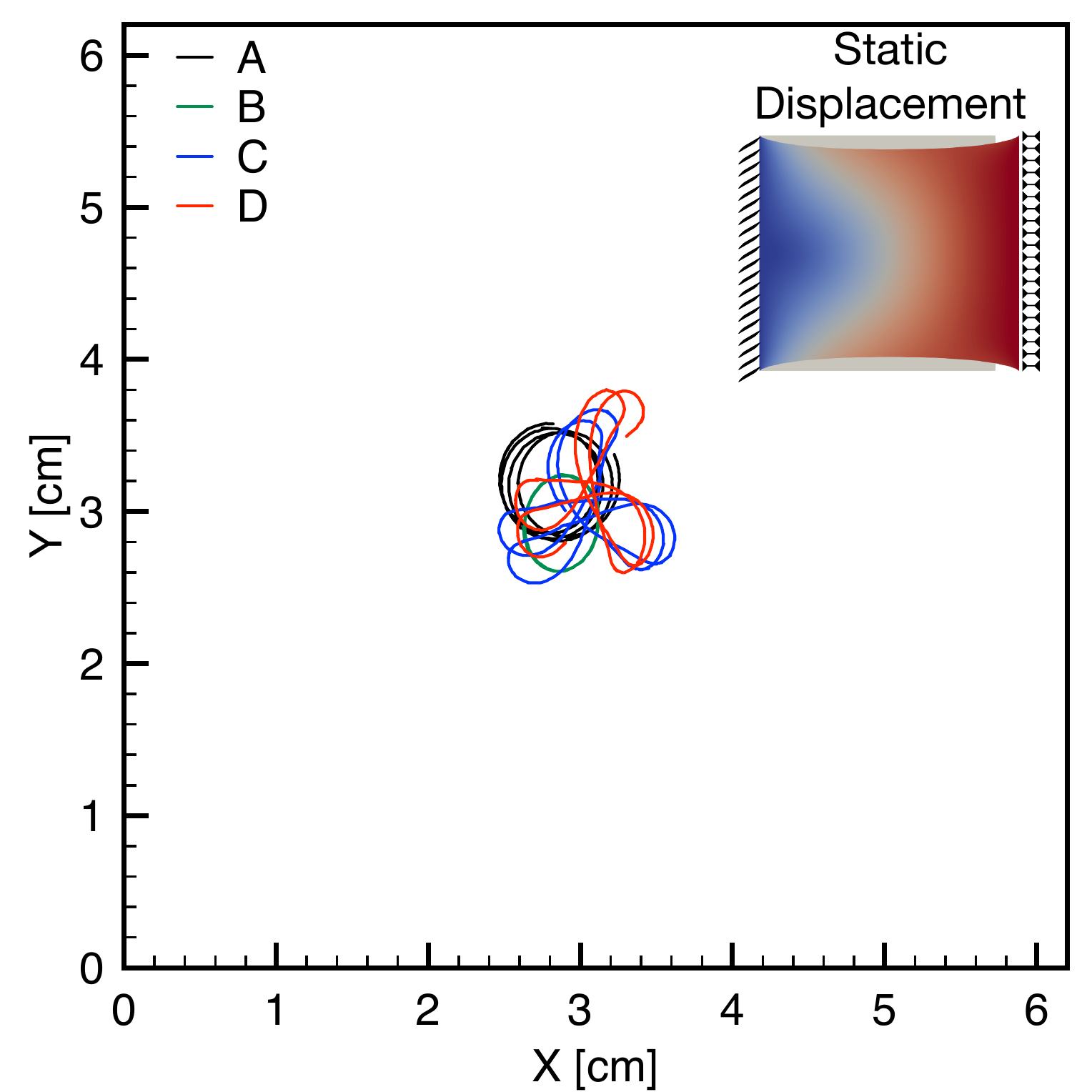}\label{fig:SD}}
\subfloat[Dynamic Displacement]{\includegraphics[width=0.48\textwidth]{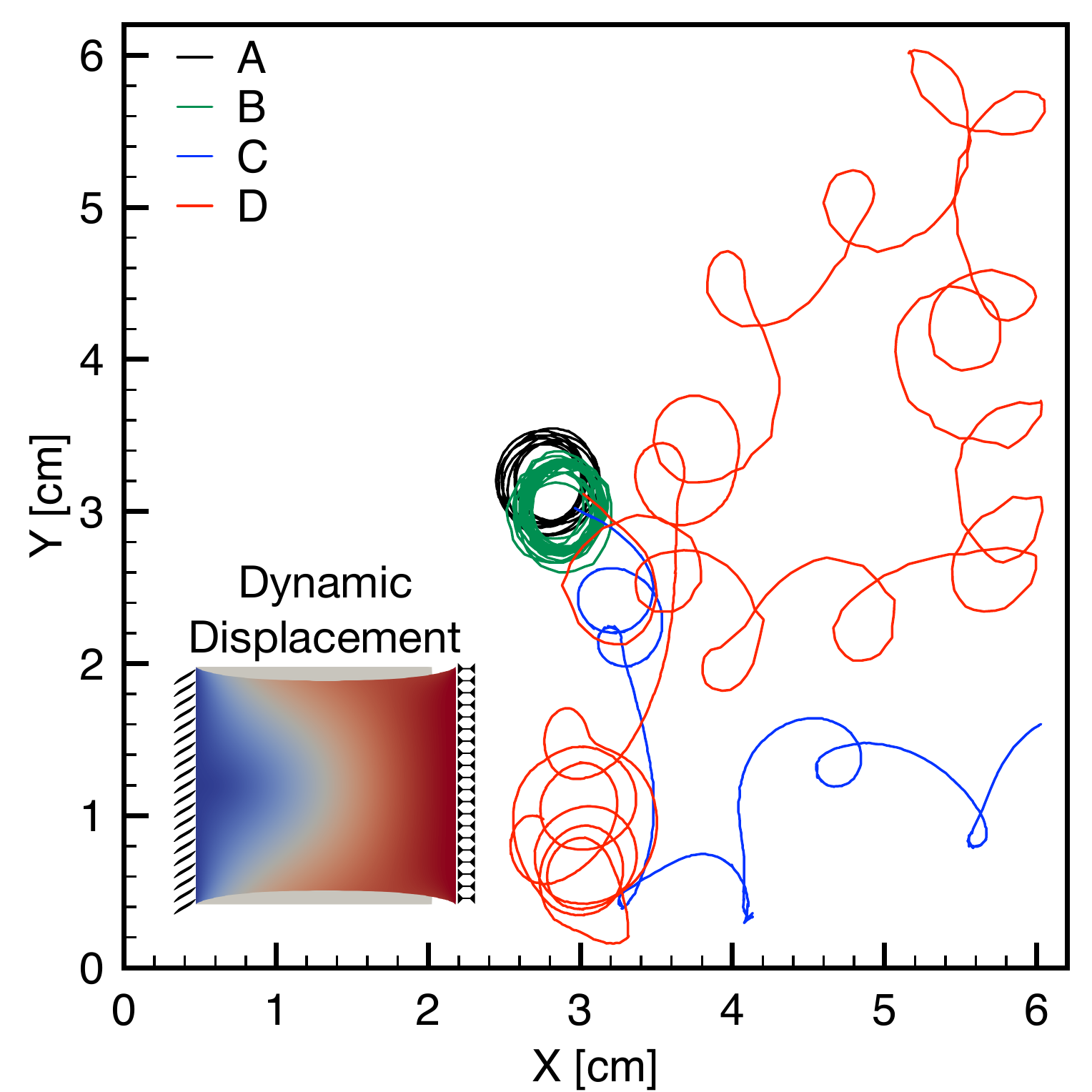}\label{fig:DD}}\\
\subfloat[Static Traction]{\includegraphics[width=0.48\textwidth]{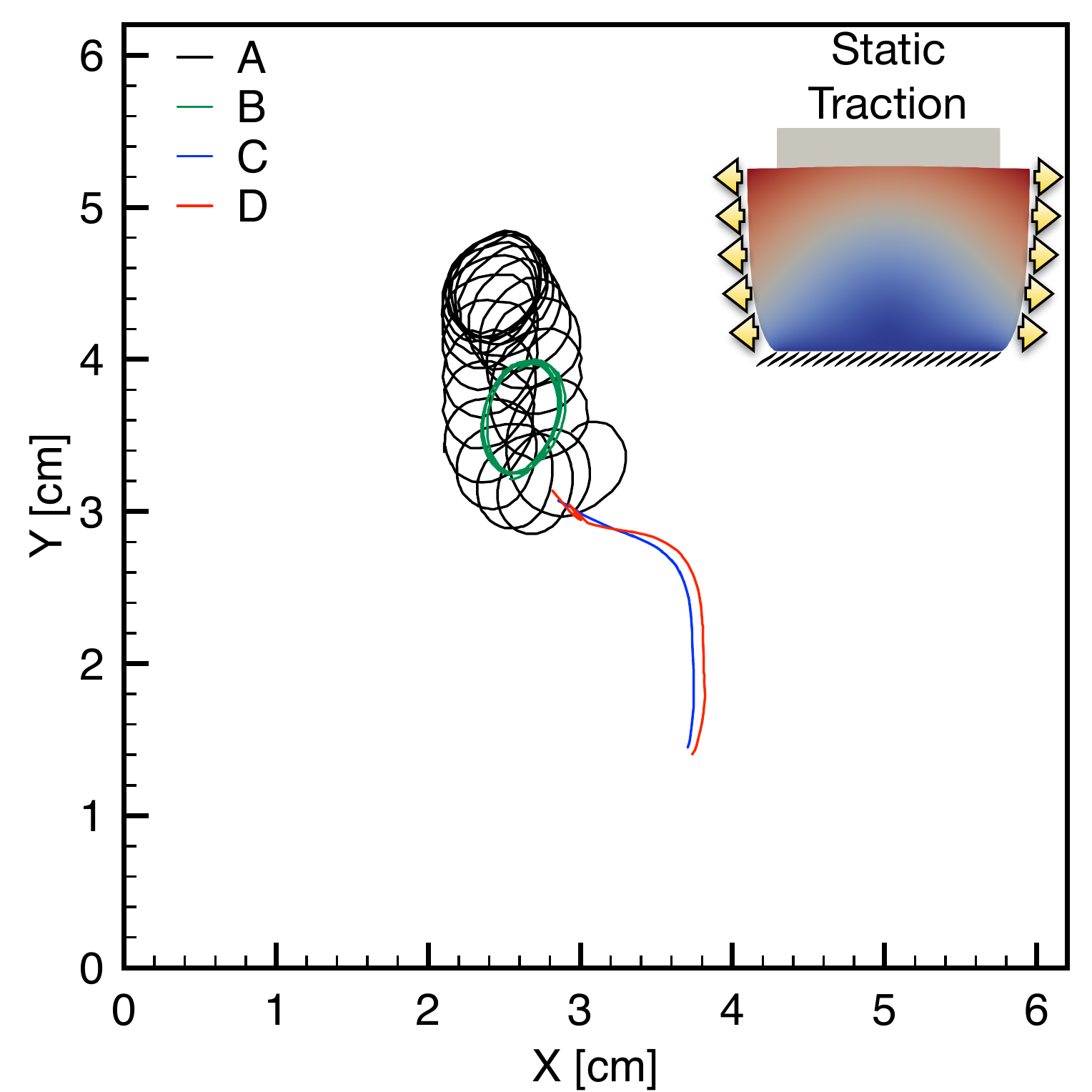}\label{fig:ST}}
\subfloat[Dynamic Traction]{\includegraphics[width=0.48\textwidth]{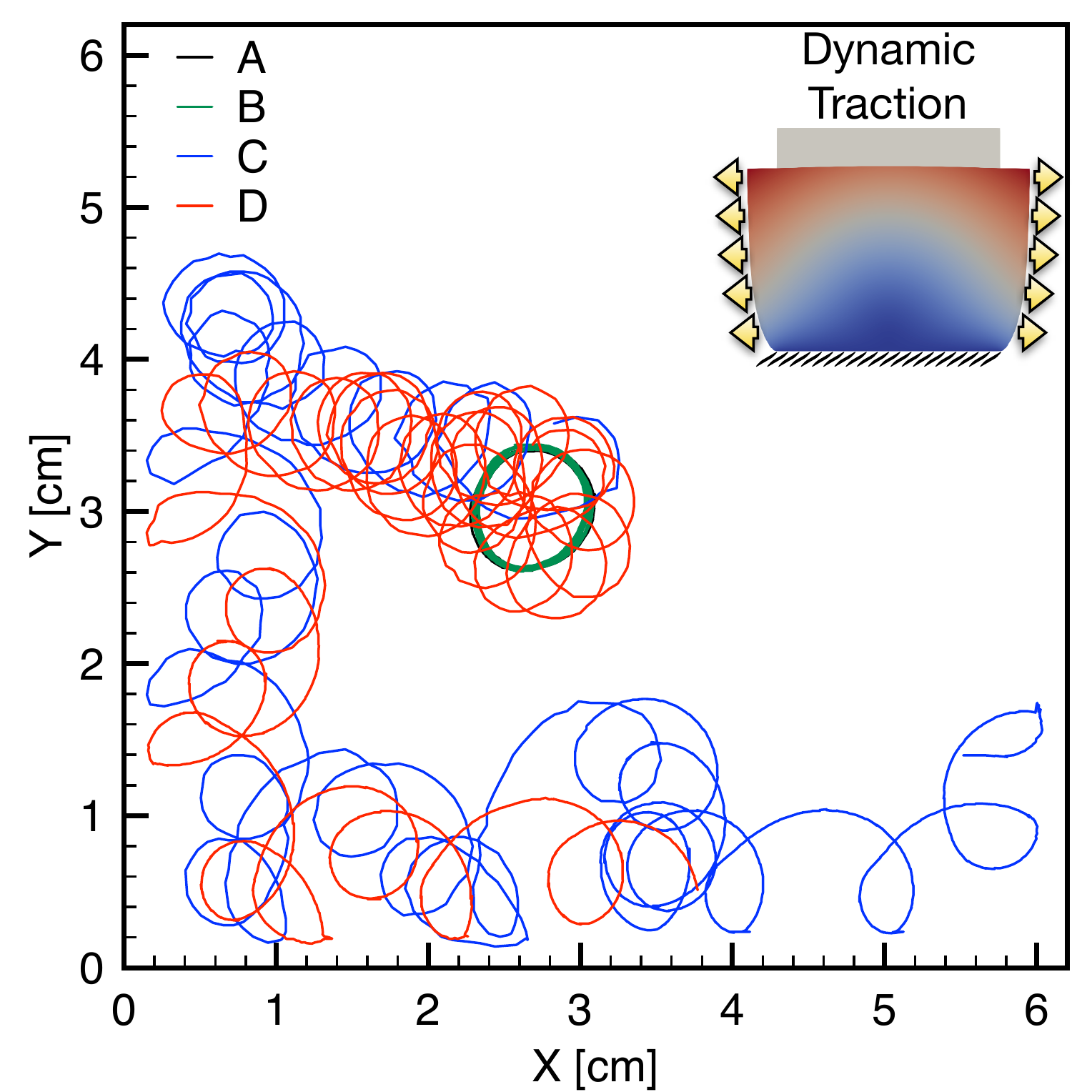}\label{fig:DT}}
\caption{Tip trajectories for four combinations of MEF
parameters $(D_1,G_s)$ (see Tab.~\ref{tab:s1_s2}), applying
static/dynamic--displacement/traction boundary conditions as indicated
in the corresponding inset. Color codes of the inset refer to the displacement
magnitude, and in the legends $D_1$ should be rescaled by $10^{-4}$.
(a) The last second of simulation is shown for the four cases with localized cores.
(b) The last three seconds of simulations are shown highlighting the differences of the meandering.
(c) Different times are shown for the four cases since for $G_s>0$ the spirals exit the domain soon after initiation.
(d) The last three seconds are shown for the case $G_s>0$ highlighting the different meandering obtained with respect to $G_s=0$.
Minor discontinuities are due to the frame resolution for post processing analysis and are not linked to the accuracy of the numerical solution.
}
\label{fig:tip}
\end{figure}

Finally, we turn to the analysis of meandering for the spiral tip 
for long run simulations 
($4\,{\rm s}$ of physical time) comparing the
four selected sets of parameters A,B,C,D in combination with
static/dynamic--displacement/traction boundary conditions. In
particular, we initiate the spiral wave via the S1-S2 stimulation
protocol as discussed in the previous section, in absence of any
mechanical loading such to start from the same initial conditions for
each selected case.  After spiral onset and stabilization 
(namely, for $t>t_2=250\,{\rm ms}$), we apply
the following four different loadings:
\begin{itemize}
\item Static displacement: uniaxial displacement
$\tilde{\pmb{\varphi}} = (0.1 L, 0)^T$ applied on the right boundary
while keeping the left one clamped (Fig.~\ref{fig:SD}).
\item Dynamic displacement: uniaxial time-dependent displacement
$\tilde{\pmb{\varphi}}(t) = \left[ 0.1L \sin^2(\pi/400 \, t),0 \right]^T$
applied on the right
boundary while keeping the left one clamped (Fig.~\ref{fig:DD}).
\item Static traction: uniaxial sigmoidal time-dependent force
$\tilde{t}_i(t) = t_{\max} \left[1.0-\exp(-(t-t_2)/5) \right]$
applied on the
left and right boundaries while keeping the bottom side clamped
(Fig.~\ref{fig:ST}).
\item Dynamic traction: uniaxial time-dependent force
$\tilde{t}_i(t) = t_{\max} \sin^2(\pi/400 \, t)$
applied on the left and right boundaries while keeping the bottom side
clamped (Fig.~\ref{fig:DT}).
\end{itemize} 

For each mechanical loading, panels in
Fig.~\ref{fig:tip} show the trajectories of the spiral tip for the
four MEF parameters combinations. Two important aspects are worthy of attention.

First, for each
combination of the mechanical loading, the presence of the
stress-assisted conductivity $D_1$ tends to stabilize the meandering
(see black and green traces). This behaviour is particularly evident
in Fig.~\ref{fig:ST} where the combination $D_1=-0.75\cdot 10^{-4},
G_s=0$ results into a localized core, while the case $D_1=0, G_s=0$
presents a circular, but slightly drifting core. Consequently, local
stress-based heterogeneities appear in the medium when $D_1$ is
different from zero, leading to pinning-like phenomena also observed
in~\cite{cherry:2008,cherubini:2012,steinbock:2012,liu:2013}.
Moreover, these conditions are associated
with an ellipsoidal shape of the core underlying the effective
anisotropy induced by the stress-assisted coupling. All these
observations agree with the conclusions from the extended analysis
conducted on the chosen AP model in the original work from
\cite{fenton:1998}.

Secondly, when also {SAC} is present, the spiral
meandering is unpredictable and strongly dependent on the applied
boundary conditions (see blue and red traces). In this scenario, it is
interesting to note that static loading induces a simple meandering
which eventually pushes the spiral wave out from the domain (see
Fig.~\ref{fig:ST}), whereas dynamic conditions dictate a chaotic
behavior that makes the spiral either to explore the whole domain, or
to exit it. These patterns seem to be extreme conditions of
hyper-excitability not expected in a two-dimensional isotropic 
medium~\citep{fenton:1998a,fenton:2002}.

\begin{figure}[h!]
\centering
\subfloat[]{\includegraphics[width=0.83\textwidth]{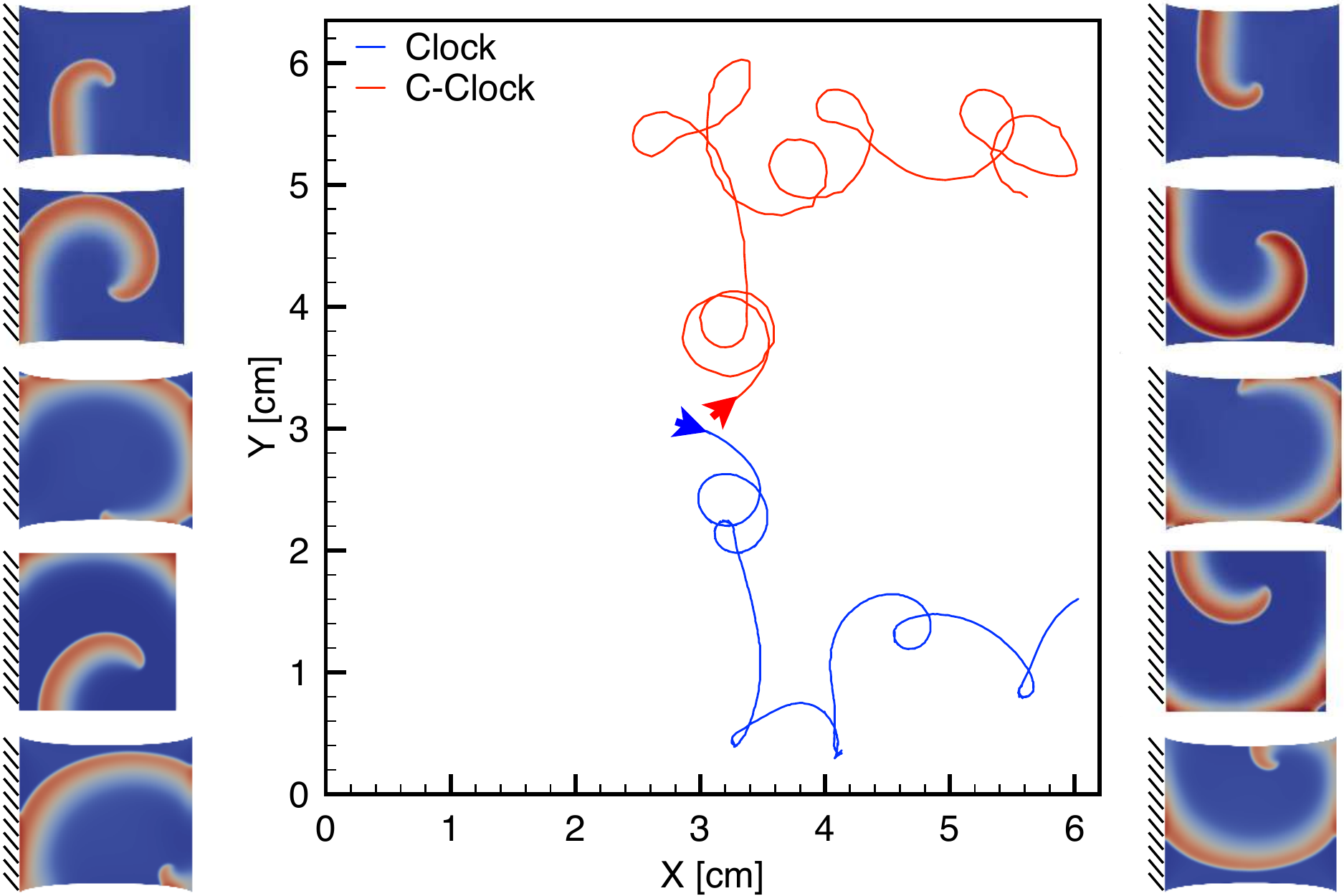}}\\
\subfloat[]{\includegraphics[width=0.83\textwidth]{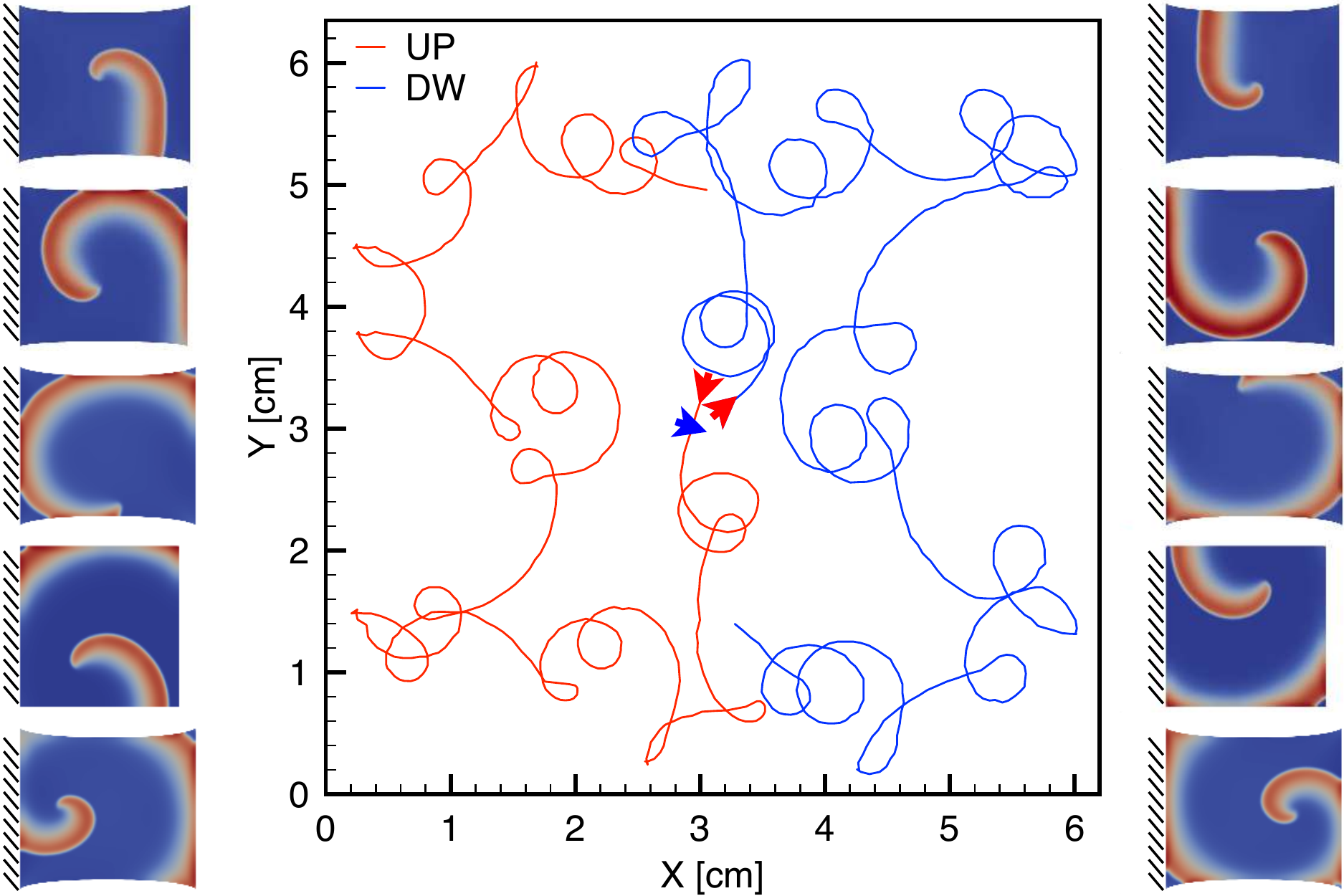}}
\caption{(a) Clockwise (blue) and counterclockwise (red) tip trajectories 
obtained in a dynamic uniaxially stretched case 
with MEF parameters $D_1=0, G_s=0.125$ and 
initiated via the S1-S2 stimulation protocol.
(b) Counterclockwise spiral initiation from top (red) or bottom (blue) boundary.
Side panels show progressive spiral frames for the two cases.}
\label{fig:tipD}
\end{figure}

Finally, we highlight the symmetry of the observed behavior according
to the clockwise or counterclockwise rotation of the spiral. This
particular analysis is provided in Fig.~\ref{fig:tipD} and further
links the excitation dynamics to the mechanical features. 
The different traces refer to the spiral core meandering
observed for a dynamic uniaxially stretched case with MEF parameters
$D_1=0, G_s=0.125$ and initiated via the S1-S2 stimulation protocol:
case (a) compares a clockwise and counterclockwise spiral propagation;
case (b) shows a counterclockwise spiral core initiated from the top (red)
and bottom (blue) case.
Corresponding sequences are also shown as side panels.
This result is limited to the simplified nature of the domain adopted, i.e.,
2D isotropic. A more realistic computational domain, embedding fiber directionality
and tissue thickness, would show more involved dynamics in a complex
spatiotemporal and clinical relevant perspective.

\section{Conclusion}\label{sec:conclusion}
We have advanced a minimal model for the electromechanics of cardiac
tissue, where the mechano-electrical feedback is incorporated through
two competing mechanisms: the stretch-activated currents commonly
found in the literature, and the stress-assisted diffusion (or
stress-assisted conductivity) recently proposed by \cite{cherubini:2017}. 
Both the electrophysiology and the
mechanical response adopt a phenomenological simplified description, but a preliminary
validation is provided through a set of numerical simulations that
agree qualitatively with a set of experimental data for pig right
ventricle.

The implications of the intensity and degree of
nonlinearity assumed for the stress-assisted diffusion effect are
studied from the viewpoint of changes in the conduction velocity and
the dynamics of spiral waves in simplified 2D domains.
Multiple electrical stimulations protocols and 
non-trivial mechanical loadings have been
investigated highlighting the strong coupling due to the different MEF contributions.
The analysis supports the hypothesis that the simplistic formulation adopted
for stretch-activated currents seems to deviate from the experimental evidence,
in line with recent contributions addressing the coupled modeling of SACs
and stretch-induced myofilament calcium release at the myocyte level~\citep{mcculloch:2017}.
On the other hand, in a homogenized setting, 
the stress-assisted diffusion formulation produces a series
of interesting phenomena that qualitatively match heterogeneities and
anisotropies observed during mechanical stretching of 
pig right ventricle via fluorescence optical mapping.

Limitations of the present work are partially linked to the 
phenomenological approach
adopted to describe the complex multiscale mechanisms intrinsic in the cardiac tissue
and partially due to the simplified computational domain.
In this regards, we aim at investigating more reliable stretch-activated current 
formulations leading to alternans behaviors~\citep{galice:2016} within
a multiscale mechanobiology perspective~\citep{nava:2016,staaland:2016,cyron:2017a} 
and tacking into account
the intracellular calcium cycling influenced by mechanical stretch, because
all these effects have been proposed as concurring mechanisms of 
arrhythmogenesis within the heart.
From the mechanical point of view, we mention as main limitation the adoption of 
a simplified isotropic hyperelastic material model which can be generalized to more complex 
and reliable formulations. This will include, for example, active strain anisotropies, 
muscular and collagen fiber distributions in an orthotropic mechanical framework 
that the authors have been extensively developing during the 
last decade~\citep{cherubini:2008,nobile:2012,gizzi:2015,pandolfi:2016,gizzi:2016a,gizzi:2018a}.
Such a generalization will maintain the nature of the present theoretical framework 
in terms of MEF competitive effects.
In this line, we also aim to generalize our theoretical and computational approach towards 
intrinsic multiscale and multiphysics mechano-transduction problems, e.g. the uterine
smooth muscle activity~\citep{yochum:2017,young:2016} or 
the intestine biomechanics activity~\citep{aydin:2017,pandolfi:2017b}
by implying the usage of network approaches~\citep{giuliani:2014,robson:2018}
and data assimilation procedures~\citep{barone:2017}.
In addition, the investigation of the complex spatiotemporal dynamics, chaos control 
and multiphysics couplings in excitable systems (see e.g. \cite{horning:2017,christoph:2018}) 
can be emphasized within the proposed electromechanical framework
by using realistic three-dimensional cardiac structures~\citep{CNM:CNM1494}. We also 
mention implications of the proposed models in the mathematical study of general stress-assisted 
diffusion problems, as recently carried out in \cite{gatica:2018}. 
Finally, we hope that the present contribution may open new experimental studies
to translate the complex MEF phenomena into the clinical practice~\citep{taggart:2017,MEIJBORG2017356} 
identifying novel risk indices for cardiac arrhythmias~\citep{gizzi:2017b}.

\section*{Acknowledgments}
This work has been supported by the Italian National Group of Mathematical Physics GNFM-INdAM; by the International Center for Relativistic Astrophysics Network ICRANet; by the London Mathematical Society through its Grant Scheme 4; and by the 
EPSRC through the Research Grant EP/R00207X/1.

\small 
\bibliographystyle{siam} 
\bibliography{frontiers}

\providecommand{\noopsort}[1]{}\providecommand{\singleletter}[1]{#1}%
\begin{thebibliography}{10}

\bibitem{aifantis:1980}
{\sc E.~C. Aifantis}, {\em On the problem of diffusion in solids}, Acta
  Mechanica, 37 (1980), pp.~265--296.

\bibitem{fenics}
{\sc M.~S. Aln{\ae}s, J.~Blechta, J.~Hake, A.~Johansson, B.~Kehlet, A.~Logg,
  C.~Richardson, J.~Ring, M.~E. Rognes, and G.~N. Wells}, {\em The {FE}ni{CS}
  project version 1.5}, Archive of Numerical Software, 3 (2015), pp.~9--23.

\bibitem{ambrosi:2012}
{\sc D.~Ambrosi and S.~Pezzuto}, {\em Active stress vs. active strain in
  mechanobiology: constitutive issues}, Journal of Elasticity, 107 (2012),
  pp.~199--212.

\bibitem{augustin:2016}
{\sc C.~M. Augustin, A.~Neic, M.~Liebmann, A.~J. Prassl, S.~A. Niederer,
  G.~Haase, and G.~Plank}, {\em Anatomically accurate high resolution modeling
  of human whole heart electromechanics: A strongly scalable algebraic
  multigrid solver method for nonlinear deformation}, Journal of Computational
  Physics, 305 (2016), pp.~622--646.

\bibitem{aydin:2017}
{\sc R.~C. Aydin, S.~Brandstaeter, F.~A. Braeu, M.~Steigenberger, R.~P. Marcus,
  K.~Nikolaou, M.~Notohamiprodjo, and C.~J. Cyron}, {\em Experimental
  characterization of the biaxial mechanical properties of porcine gastric
  tissue}, Journal of the Mechanical Behavior of Biomedical Materials, 74
  (2017), pp.~499--506.

\bibitem{barone:2017}
{\sc A.~Barone, F.~H. Fenton, and A.~Veneziani}, {\em Numerical sensitivity
  analysis of a variational data assimilation procedure for cardiac
  conductivities}, Chaos: An Interdisciplinary Journal of Nonlinear Science, 27
  (2017), p.~093930.

\bibitem{bini:2010}
{\sc D.~Bini, C.~Cherubini, S.~Filippi, A.~Gizzi, and P.~E. Ricci}, {\em On
  spiral waves arising in natural systems}, Communications in Computational
  Physics, 8 (2010), p.~610.

\bibitem{cabo:2014}
{\sc C.~Cabo}, {\em Dynamics of propagation of premature impulses in
  structurally remodeled infarcted myocardium: a computational analysis},
  Frontiers in Physiology, 5 (2014), p.~483.

\bibitem{chen:2014}
{\sc J.-X. Chen, L.~Peng, Q.~Zheng, Y.-H. Zhao, and H.-P. Ying}, {\em
  Influences of periodic mechanical deformation on pinned spiral waves}, Chaos:
  An Interdisciplinary Journal of Nonlinear Science, 24 (2014), p.~033103.

\bibitem{cherry:2008}
{\sc E.~M. Cherry and F.~H. Fenton}, {\em Visualization of spiral and scroll
  waves in simulated and experimen- tal cardiac tissue}, New Journal of
  Physics, 10 (2008), p.~125016.

\bibitem{cherubini:2012}
{\sc C.~Cherubini, S.~Filippi, and A.~Gizzi}, {\em Electroelastic unpinning of
  rotating vortices in biological excitable media}, Physical Review E, 85
  (2012), p.~031915.

\bibitem{cherubini:2017}
{\sc C.~Cherubini, S.~Filippi, A.~Gizzi, and R.~Ruiz-Baier}, {\em A note on
  stress-driven anisotropic diffusion and its role in active deformable media},
  Journal of Theoretical Biology, 430 (2017), pp.~221--228.

\bibitem{cherubini:2008}
{\sc C.~Cherubini, S.~Filippi, P.~Nardinocchi, and L.~Teresi}, {\em An
  electromechanical model of cardiac tissue: Constitutive issues and
  electrophysiological effects}, Progress in Biophysics and Molecular Biology,
  97 (2008), pp.~562--573.

\bibitem{christoph:2018}
{\sc J.~Christoph, M.~Chebbok, C.~Richter, J.~Schr{\"o}der-Schetelig,
  P.~Bittihn, S.~Stein, I.~Uzelac, F.~H. Fenton, G.~Hasenfu{\ss}, J.~Gilmour,
  R.~F., and S.~Luther}, {\em Electromechanical vortex filaments during cardiac
  fibrillation}, Nature, 555 (2018), p.~667.

\bibitem{hurtado:2017a}
{\sc F.~S. Costabal, F.~A. Concha, D.~E. Hurtado, and E.~Kuhl}, {\em The
  importance of mechano-electrical feedback and inertia in cardiac
  electromechanics}, Computer Methods in Applied Mechanics and Engineering, 320
  (2017), pp.~352--368.

\bibitem{cyron:2017a}
{\sc C.~J. Cyron and J.~D. Humphrey}, {\em Growth and remodeling of
  load-bearing biological soft tissues}, Meccanica, 52 (2017), pp.~645--664.

\bibitem{dhein:2014}
{\sc S.~Dhein, T.~Seidel, A.~Salameh, J.~Jozwiak, A.~Hagen, M.~Kostelka,
  G.~Hindricks, and F.~W. Mohr}, {\em Remodeling of cardiac passive electrical
  properties and susceptibility to ventricular and atrial arrhythmias},
  Frontiers in Physiology, 5 (2014), p.~424.

\bibitem{dierckx:2015}
{\sc H.~Dierckx, S.~Arens, B.-W. Li, L.~D. Weise, and A.~V. Panfilov}, {\em A
  theory for spiral wave drift in reaction-diffusion-mechanics systems}, New
  Journal of Physics, 17 (2015), p.~043055.

\bibitem{fenton:2008}
{\sc F.~H. Fenton and E.~M. Cherry}, {\em Models of cardiac cell},
  Scholarpedia, 3 (2008), p.~1868.

\bibitem{fenton:2002}
{\sc F.~H. Fenton, E.~M. Cherry, H.~M. Hasting, and S.~J. Evans}, {\em Multiple
  mechanisms of spiral wave breakup in a model of cardiac electrical activity},
  Chaos, 12 (2002), pp.~852--892.

\bibitem{fenton:1998a}
{\sc F.~H. Fenton and A.~Karma}, {\em Fiber-rotation-induced vortex turbulence
  in thick myocardium}, Physical Review Letters, 81 (1998), p.~481.

\bibitem{fenton:1998}
\leavevmode\vrule height 2pt depth -1.6pt width 23pt, {\em Vortex dynamics in
  three-dimensional continuous myocardium with fiber rotation: Filament
  instability and fibrillation}, Chaos, 8 (1998), pp.~20--47.

\bibitem{fenton:2009}
{\sc F.~H. Fenton, S.~Luther, N.~F. Otani, V.~Krinsky, A.~Pumir,
  E.~Bodenschatz, and J.~Gilmour, R.~F.}, {\em Termination of atrial
  fibrillation using pulsed low-energy far-field stimulation}, Circulation, 120
  (2009), pp.~467--476.

\bibitem{galice:2016}
{\sc S.~Galice, D.~M. Bers, and D.~Sato}, {\em Stretch-activated current can
  promote or suppress cardiac alternans depending on voltage-calcium
  interaction}, Biophysical Journal, 110 (2016), pp.~2671--2677.

\bibitem{gatica:2018}
{\sc G.~N. Gatica, B.~Gomez-Vargas, and R.~Ruiz-Baier}, {\em Analysis and
  mixed-primal finite element discretisations for stress-assisted diffusion
  problems}, Computer Methods in Applied Mechanics and Engineering,  (2018),
  pp.~1--28.

\bibitem{giuliani:2014}
{\sc A.~Giuliani, S.~Filippi, and M.~Bertolaso}, {\em Why network approach can
  promote a new way of thinking in biology}, Frontiers in Genetics, 5 (2014).

\bibitem{gizzi:2013}
{\sc A.~Gizzi, E.~M. Cherry, J.~Gilmour, R.~F., S.~Luther, S.~Filippi, and
  F.~H. Fenton}, {\em Effects of pacing site and stimulation history on
  alternans dynamics and the development of complex spatiotemporal patterns in
  cardiac tissue}, Frontiers in Physiology, 4 (2013), p.~71.

\bibitem{gizzi:2015}
{\sc A.~Gizzi, C.~Cherubini, S.~Filippi, and A.~Pandolfi}, {\em Theoretical and
  numerical modeling of nonlinear electromechanics with applications to
  biological active media}, Communications in Computational Physics, 17 (2015),
  pp.~93--126.

\bibitem{gizzi:2017b}
{\sc A.~Gizzi, A.~Loppini, E.~M. Cherry, C.~Cherubini, F.~H. Fenton, and
  S.~Filippi}, {\em Multi-band decomposition analysis: {A}pplication to cardiac
  alternans as a function of temperature}, Physiological Measurements, 38
  (2017), pp.~833--847.

\bibitem{gizzi:2016a}
{\sc A.~Gizzi, A.~Pandolfi, and M.~Vasta}, {\em Statistical characterization of
  the anisotropic strain energy in soft materials with distributed fibers},
  Mechanics of Materials, 92 (2016), pp.~119--138.

\bibitem{gizzi:2018a}
\leavevmode\vrule height 2pt depth -1.6pt width 23pt, {\em A generalized
  statistical approach for modeling fiber-reinforced materials}, Journal of
  Engineering Mathematics, 109 (2018), pp.~211--226.

\bibitem{horning:2012}
{\sc M.~H\"orning}, {\em Termination of pinned vortices by high-frequency wave
  trains in heartlike excitable media with anisotropic fiber orientation},
  Physical Review E, 86 (2012), p.~031912.

\bibitem{horning:2017}
{\sc M.~H\"orning, F.~Blanchard, A.~Isomura, and K.~Yoshikawa}, {\em Dynamics
  of spatiotemporal line defects and chaos control in complex excitable
  systems}, Scientific Reports, 7 (2017), p.~7757.

\bibitem{hunter:1997}
{\sc P.~J. Hunter, M.~P. Nash, and G.~B. Sands}, {\em Computational
  electromechanics of the heart}, Computational Biology of the Heart, 12
  (1997), pp.~347--407.

\bibitem{hurtado:2016}
{\sc D.~E. Hurtado, S.~Castro, and A.~Gizzi}, {\em Computational modeling of
  non-linear diffusion in cardiac electrophysiology: A novel porous-medium
  approach}, Computer Methods in Applied Mechanics and Engineering, 300 (2016),
  pp.~70--83.

\bibitem{trayanova:2010}
{\sc X.~Jie, V.~Gurev, and N.~A. Trayanova}, {\em Mechanisms of mechanically
  induced spontaneous arrhythmias in acute regional ischemia}, Circulation
  Research, 106 (2010), pp.~185--192.

\bibitem{steinbock:2012}
{\sc Z.~A. Jimenez and O.~Steinbock}, {\em Scroll wave filaments self-wrap
  around unexcitable heterogeneities}, Physical Review E, 86 (2012), p.~036205.

\bibitem{karma:2013}
{\sc A.~Karma}, {\em Physics of cardiac arrhythmogenesis}, Annual Review of
  Condensed Matter Physics, 4 (2013), pp.~313---337.

\bibitem{panfilov:2010}
{\sc R.~H. Keldermann, M.~P. Nash, H.~Gelderblom, V.~Y. Wang, and A.~V.
  Panfilov}, {\em Electromechanical wavebreak in a model of the human left
  ventricle}, American Journal of Physiology-Heart and Circulatory Physiology,
  299 (2010), pp.~H134--H143.

\bibitem{kleber:2014}
{\sc A.~G. Kleber and J.~E. Saffitz}, {\em Role of the intercalated disc in
  cardiac propagation and arrhythmogenesis}, Frontiers in Physiology, 5 (2014),
  p.~404.

\bibitem{CNM:CNM1494}
{\sc P.~Lafortune, R.~Ar{\'\i}s, M.~V{\'a}zquez, and G.~Houzeaux}, {\em Coupled
  electromechanical model of the heart: Parallel finite element formulation},
  International Journal for Numerical Methods in Biomedical Engineering, 28
  (2012), pp.~72--86.

\bibitem{land:2015}
{\sc S.~Land and et. al.}, {\em Verification of cardiac mechanics software:
  benchmark problems and solutions for testing active and passive material
  behaviour}, {Proc. R. Soc. Lond. A}, 471 (2016), p.~20150641.

\bibitem{land:2017}
{\sc S.~Land, S.~J. Park-Holohan, N.~P. Smith, C.~G. Dos~Remedios, J.~C.
  Kentish, and S.~A. Niederer}, {\em A model of cardiac contraction based on
  novel measurements of tension development in human cardiomyocytes}, Journal
  of Molecular and Cellular Cardiology, 106 (2017), pp.~68--83.

\bibitem{trayanova:2004}
{\sc W.~Li, P.~Kohl, and N.~A. Trayanova}, {\em Induction of ventricular
  arrhythmias following mechanical impact: a simulation study in 3{D}}, Journal
  of Molecular Histology, 35 (2004), pp.~679--686.

\bibitem{liu:2013}
{\sc T.~B. Liu, J.~Ma, Q.~Zhao, and J.~Tang}, {\em Force exerted on the spiral
  tip by the heterogeneity in an excitable medium}, Europhysics Letters, 104
  (2013), p.~58005.

\bibitem{MEIJBORG2017356}
{\sc V.~M.~F. Meijborg, C.~N.~W. Belterman, J.~M.~T. de~Bakker, R.~Coronel, and
  C.~E. Conrath}, {\em Mechano-electric coupling, heterogeneity in
  repolarization and the electrocardiographic t-wave}, Progress in Biophysics
  and Molecular Biology, 130 (2017), pp.~356--364.

\bibitem{nava:2016}
{\sc M.~M. Nava, R.~Fedele, and M.~T. Raimondi}, {\em Computational prediction
  of strain-dependent diffusion of transcription factors through the cell
  nucleus}, Biomechanics and Modeling in Mechanobiology, 15 (2016),
  pp.~983--993.

\bibitem{nobile:2012}
{\sc F.~Nobile, R.~Ruiz-Baier, and A.~Quarteroni}, {\em An active strain
  electromechanical model for cardiac tissue}, International Journal for
  Numerical Methods in Biomedical Engineering, 28 (2012), pp.~52--71.

\bibitem{taggart:2017}
{\sc M.~Orini, A.~Nanda, M.~Yates, C.~Di~Salvo, N.~Roberts, P.~D. Lambiasea,
  and P.~Taggart}, {\em Mechano-electrical feedback in the clinical setting:
  Current perspectives}, Progress in Biophysics and Molecular Biology, 130
  (2017), pp.~365--375.

\bibitem{pandolfi:2016}
{\sc A.~Pandolfi, A.~Gizzi, and M.~Vasta}, {\em Coupled electro-mechanical
  models of fiber-distributed active tissues}, J. Biomech., 49 (2016),
  pp.~2436--2444.

\bibitem{pandolfi:2017b}
\leavevmode\vrule height 2pt depth -1.6pt width 23pt, {\em
  Visco-electro-elastic models of fiber-distributed active tissues}, Meccanica,
  52 (2017), p.~3399.

\bibitem{panfilov:2005}
{\sc A.~V. Panfilov and R.~H. Keldermann}, {\em Self-organized pacemakers in a
  coupled reaction-diffusion-mechanics system}, Physical Review Letters, 95
  (2005), p.~258104.

\bibitem{pullan:2005}
{\sc A.~J. Pullan, L.~K. Cheng, and M.~L. Buist}, {\em Mathematically Modelling
  the Electrical Activity of the Heart: From Cell to Body Surface and Back
  Again}, World Scientific, 2005.

\bibitem{quarteroni:2017}
{\sc A.~Quarteroni, T.~Lassila, S.~Rossi, and R.~Ruiz~Baier}, {\em Integrated
  heart -- coupled multiscale and multiphysics models for the simulation of the
  cardiac function}, Computer Methods in Applied Mechanics and Engineering, 314
  (2017), pp.~345--407.

\bibitem{qv94}
{\sc A.~Quarteroni and A.~Valli}, {\em Numerical approximation of partial
  differential equations}, vol.~23 of Springer Series in Computational
  Mathematics, Springer-Verlag, Berlin, 1994.

\bibitem{quinn:2016}
{\sc T.~A. Quinn and P.~Kohl}, {\em Rabbit models of cardiac mechano-electric
  and mechano-mechanical coupling}, Progress in Biophysics and Molecular
  Biology, 121 (2016), pp.~110--122.

\bibitem{quinn:2014}
{\sc T.~A. Quinn, P.~Kohl, and U.~Ravens}, {\em Cardiac mechano-electric
  coupling research: Fifty years of progress and scientific innovation},
  Progress in Biophysics and Molecular Biology, 115 (2014), pp.~71--75.

\bibitem{ravelli:2003}
{\sc F.~Ravelli}, {\em Mechano-electric feedback and atrial fibrillation},
  Progress in Biophysics and Molecular Biology, 82 (2003), pp.~137--149.

\bibitem{robson:2018}
{\sc J.~Robson, P.~Aram, M.~P. Nash, C.~P. Bradley, M.~Hayward, D.~J. Paterson,
  P.~Taggart, R.~H. Clayton, and V.~Kadirkamanathan}, {\em Spatio-temporal
  organization during ventricular fibrillation in the human heart}, Annals of
  Biomedical Engineering,  (2018).

\bibitem{rossi:2014}
{\sc S.~Rossi, T.~Lassila, R.~Ruiz-Baier, A.~Sequeira, and A.~Quarteroni}, {\em
  Thermodynamically consistent orthotropic activation model capturing
  ventricular systolic wall thickening in cardiac electromechanics}, European
  Journal of Mechanics: A/Solids, 48 (2014), pp.~129--142.

\bibitem{ruiz:2015}
{\sc R.~Ruiz-Baier}, {\em Primal-mixed formulations for reaction-diffusion
  systems on deforming domains}, Journal of Computational Physics, 299 (2015),
  pp.~320--338.

\bibitem{salameh:2013}
{\sc A.~Salamhe and S.~Dhein}, {\em Effects of mechanical forces and stretch on
  intercellular gap junction coupling}, Biochimica et Biophysica Acta (BBA) -
  Biomembranes, 1828 (2013), pp.~147--156.

\bibitem{schonleitner:2017}
{\sc P.~Sch\"onleitner, U.~Schotten, and G.~Antoons}, {\em Mechanosensitivity
  of microdomain calcium signalling in the heart}, Progress in Biophysics and
  Molecular Biology, 130 (2017), pp.~1--14.

\bibitem{spencer:1980}
{\sc A.~J.~M. Spencer}, {\em Continuum Mechanics}, Longman Group Ltd, London,
  1989.

\bibitem{staaland:2016}
{\sc J.~St{\aa}lhand, R.~M. McMeeking, and G.~A. Holzapfel}, {\em On the
  thermodynamics of smooth muscle contraction}, Journal of the Mechanics and
  Physics of Solids, 94 (2016), pp.~490--503.

\bibitem{tadmor:2012}
{\sc E.~B. Tadmor, R.~E. Miller, and R.~S. Elliot}, {\em Continuum mechanics
  and thermodynamics: {F}rom fundamental concepts to governing equations},
  Cambridge University Press., 2012.

\bibitem{mcculloch:2017}
{\sc V.~Timmermann, L.~A. Dejgaard, K.~H. Haugaa, A.~G. Edwards, J.~Sundnes,
  A.~D. McCulloch, and S.~T. Wall}, {\em An integrative appraisal of
  mechano-electric feedback mechanisms in the heart}, Progress in Biophysics
  and Molecular Biology, 130 (2017), pp.~404--417.

\bibitem{trayanova:2006}
{\sc N.~A. Trayanova}, {\em Defibrillation of the heart: insights into
  mechanisms from modelling studies}, Experimental Physiology, 91 (2006),
  pp.~323--337.

\bibitem{trayanova:2011a}
{\sc N.~A. Trayanova and J.~J. Rice}, {\em Cardiac electromechanical models:
  from cell to organ}, Frontiers in Physiology, 2 (2011), p.~43.

\bibitem{fenton:2017a}
{\sc I.~Uzelac, Y.~C. Ji, D.~Hornung, J.~Schr\"oder-Scheteling, S.~Luther,
  R.~A. Gray, E.~M. Cherry, and F.~H. Fenton}, {\em Simultaneous quantification
  of spatially discordant alternans in voltage and intracellular calcium in
  langendorff-perfused rabbit hearts and inconsistencies with models of cardiac
  action potentials and ca transients}, Frontiers in Physiology, 8 (2017),
  p.~819.

\bibitem{yochum:2017}
{\sc M.~Yochum, J.~Lafor\^et, and C.~Marque}, {\em Multi-scale and
  multi-physics model of the uterine smooth muscle with mechanotransduction},
  Computers in Biology and Medicine, 93 (2017), pp.~17--30.

\bibitem{young:2016}
{\sc R.~C. Young}, {\em Mechanotransduction mechanisms for coordinating uterine
  contractions in human labour}, Reproduction, 152 (2016), pp.~R51--61.

\end{thebibliography}

\end{document}